\documentclass[sn-nature]{sn-jnl}%

\usepackage{graphicx}%
\usepackage{multirow}%
\usepackage{amsmath,amssymb,amsfonts}%
\usepackage{amsthm}%
\usepackage{mathrsfs}%
\usepackage[title]{appendix}%
\usepackage{xcolor}%
\usepackage{textcomp}%
\usepackage{manyfoot}%
\usepackage{booktabs}%
\usepackage{algorithm}%
\usepackage{algorithmicx}%
\usepackage{algpseudocode}%
\usepackage{listings}%
\usepackage{appendix}

\newcommand{\MuST}{MuST}
\usepackage{soul, color, xcolor}
\usepackage{bm}
\usepackage{lscape}

\usepackage{multirow}
\usepackage{booktabs}
\usepackage{hyperref}

\theoremstyle{thmstyleone}%

\theoremstyle{thmstyletwo}%

\theoremstyle{thmstylethree}%

\begin{document}

\title{Maximizing Latent Capacity of Spatial Transcriptomics Data}

\author[1]{\fnm{Zelin Zang}}
\equalcont{These authors contributed equally to this work.}

\author[1]{\fnm{Liangyu Li}}
\equalcont{These authors contributed equally to this work.}

\author[1]{\fnm{Yongjie Xu}}
\equalcont{These authors contributed equally to this work.}

\author[1]{\fnm{Chenrui Duan}}

\author[2]{\fnm{Kai Wang}}
\author[2]{\fnm{Yang You}}
\author[3]{\fnm{Yi Sun}}
\author*[1]{\fnm{Stan Z. Li}}\email{Stan.Z.Li@westlake.edu.cn}

\affil[1]{\orgdiv{AI Lab, Research Center for Industries of the Future,} \orgname{Westlake University}}

\affil[2]{National University of Singapore}
\affil[3]{Key Laboratory of Growth Regulation and Translational Research of Zhejiang Province, School of Life Sciences, Westlake University}

\abstract{
Spatial transcriptomics (ST) technologies have revolutionized the study of gene expression patterns in tissues by providing multimodality data in transcriptomic, spatial, and morphological, offering opportunities for understanding tissue biology beyond transcriptomics. 
However, we identify the modality bias phenomenon in ST data species, i.e., the inconsistent contribution of different modalities to the labels leads to a tendency for the analysis methods to retain the information of the dominant modality. How to mitigate the adverse effects of modality bias to satisfy various downstream tasks remains a fundamental challenge. This paper introduces Multiple-modality Structure Transformation, named {\MuST}, a novel methodology to tackle the challenge. {\MuST} integrates the multi-modality information contained in the ST data effectively into a uniform latent space to provide a foundation for all the downstream tasks. It learns intrinsic local structures by topology discovery strategy and topology fusion loss function to solve the inconsistencies among different modalities. Thus, these topology-based and deep learning techniques provide a solid foundation for a variety of analytical tasks while coordinating different modalities. The effectiveness of {\MuST} is assessed by performance metrics and biological significance. The results show that it outperforms existing state-of-the-art methods with clear advantages in the precision of identifying and preserving structures of tissues and biomarkers. {\MuST} offers a versatile toolkit for the intricate analysis of complex biological systems. \\
}

\maketitle

\section{Introduction}
While single-cell RNA sequencing~(scRNA-seq) captures gene expression profiles of organisms at the single-cell resolution~\cite{kolodziejczyk2015technology,saliba2014single,luecken2019current,erhard2022time}, more recent advances in spatial transcriptomics (ST) technology provide additional multiple modality information about the spatial and morphology patterns of tissues~\cite{satija2015spatial, williams2022introduction}. 
The breakthrough technologies for ST, such as 10x Visium~\cite{hudson2022localization}, Slideseq~\cite{rodriques2019slide}, SlideseqV2~\cite{stickels2021highly}, and Stereo-seq~\cite{chen2022spatiotemporal} that measure gene expression information and graphical information in captured locations (referred to as ``spot") at a resolution of several cells or even sub-cells levels.
The enrichment has heralded potentials for understanding, interpreting, and visualizing the intricate micro-structures and underlying biological processes within tissues~\cite{staahl2016visualization,longo2021integrating, rao2021exploring, tian2023expanding}. The ST technology has been used to tackle a variety of downstream analysis tasks
including (a) spatial clustering~\cite{long2023spatially, bao2022integrative}, (b) spot visualization~\cite{dries2019giotto,xu2023structure}, (c) spatial deconvolution~\cite{kleshchevnikov2022cell2location, ma2022spatially}, (d) marker gene analysis~\cite{dumitrascu2021optimal}, and (e) spatial trajectory inference~\cite{wolf2019paga}.

We find that ST multiple modality data suffer from the modality bias phenomenon (details in the ``Results'' section). The modality bias phenomenon in ST data is similar to the concepts of those in the image and language fields~\cite{gat2020removing,guo2023modality}, which refers to the inconsistent contribution of different modality data to the label. For example, gene expression information is considered much more important than morphology~(Mor.) information~\cite{long2023spatially,dong2022deciphering,ren2022identifying}. However, for some complex tissue structures, the Mor. data can significantly improve the sharpness of the embedding and the precision of the biological analysis. Similarly, for some ST techniques that do not incorporate Mor. data, a better balance between spatial location and gene expression also improves the performance of the method.
Failure to address this modality bias will result in loss of information, thus hindering the ability to fully exploit the multiple modality information of ST data for joint analysis of multiple downstream tasks. The issue of modality bias in ST needs to be addressed.

Existing ST methods fall into two main categories: clustering-based and (dis)similarity-based techniques. Clustering-based techniques, such as those based on contrastive~learning~\cite{zong2022const, zeng2023identifying, long2023spatially} and deep embedding clustering~\cite{zong2022const, li2022cell, bao2022integrative, teng2022clustering, hu2021spagcn}, focus on preserving global information. Regardless of the severity of the modality bias, such methods discard local information, which leads to the failure of such methods in downstream tasks that require accurate local information~(such as spot visualisation and spatial deconvolution).
The (dis)similarity-based techniques, including kernel learning~\cite{pham2020stlearn, dong2022deciphering, zhang2022graph, xu2022deepst} and distance fitting~\cite{bergenstraahle2022super} methods, train models on similarities or dissimilarities. These methods use fixed weighting coefficients between different modalities to estimate the contribution in the data fusion process and neglect the important information in weaker modalities. In addition, a fixed similarity also leads to local conflicts between the spots.

\begin{figure}[ht]
    \centering
    \includegraphics[width=0.9\textwidth]{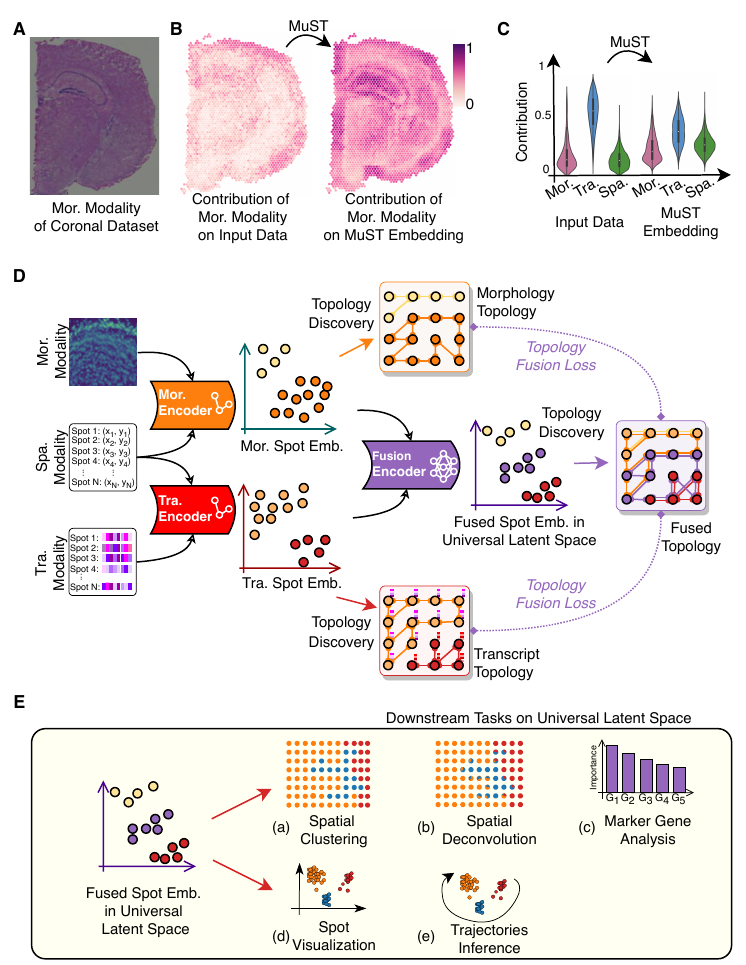}
    \caption{
    \textbf{Illustration of {\MuST}.} 
    \textbf{A} Morphology~(Mor.) modality data of the coronal dataset. 
    \textbf{B} Illustration of the modality bias phenomenon. In the case of Mor. modality, there is a significant inconsistency in the contribution of Mor. data to the final result (detected by Shapley value~\cite{winter2002shapley}). This inconsistency causes many methods to discard weaker modalities such as Mor. and Spa. resulting in a loss of information. MuST improves the retention of weaker modalities by mapping the Mor modality to the Transcriptomic~(Tra.) modality. The results of the interpretable analyses show that the phenomenon of modality bias is mitigated by the Mor. Encoder processing.
    \textbf{C} The statistical evidence of the modality bias phenomenon is mitigated.
    \textbf{D} The ST data and its topologies of Tra. modalities, Mor. modalities and Spa. modalities are fused into a universal latent feature space (universal latent space) in which the information loss problems caused by the modality bias phenomenon are mitigated and the topologies of multimodal data are well represented. Specifically, the unit of ST data is called a 'spot', and each spot contains Tra. data, Spa. data and Mor. data. MuST aims to map each spot into a vector of universal latent space in order to perform various downstream tasks. The Mor. and Tra. modality data are mapped to the single modal latent space feature by two graph encoders, with the spatial modal information used to construct the edges of the graph. The graph encoders represent each point of the ST data as a vector. The individual modal latent space features are then mapped to the universal latent space using a fusion encoder. The topology discovery operation extracts topological structures from the embedding by constructing KNN graphs (see Sec.~\ref{Graph_Construction_method}), which express the topological relationships of the spot embedding. The topology fusion loss guides the network to learn a unified representation capable of fusing multimodal data, thereby improving performance in several downstream tasks.
    \textbf{E} Multiple downstream tasks. MuST incorporates information from multiple modalities and better supports multiple downstream tasks. These include (a) spatial clustering, (b) spot visualisation, (c) spatial deconvolution, (d) marker gene analysis, and (e) spatial trajectory inference. The mutual support of multiple downstream tasks allows for a more comprehensive understanding of ST data.
    }
    \label{fig_method_overview}
\end{figure}

We introduced the Multiple-modality Structure Transformation ({\MuST}) combining augmentation-based manifold learning~\cite{zang2022deep,zang2022dlme} with graph neural networks~\cite{kipf2017semisupervised}, which creates a universal latent space that integrates multiple input modalities and facilitating in-depth analysis of ST data across various downstream tasks. 
To mitigate the adverse effects of the modality bias phenomenon, {\MuST} do not directly fuse the information of modalities, instead of measuring the modality's importance in Mor. and transcriptomic~(Tra.) spot embedding space~(Fig.~\ref{fig_method_overview}D). Specifically, we proposed a topology discovery strategy to accurate estimation global and local information of multiple modality data by extracting the topology from Mor./Tra. spot embedding space. In addition, a topology fusion loss is proposed to minimize the topological differences between all modalities and the latent space with an adaptive motility's importance estimated from the continuously updated Mor./Tra.'s spot embedding. The effectiveness of {\MuST} is evaluated comprehensively in terms of scientific indicators and biological significance by comparison with state-of-the-art methods on ST data of human and mouse tissues generated by different platforms (e.g., 10x Visium, Slide-seqV2, and Stereo-seq). On the scientific metrics, {\MuST} has a clear advantage on the geometric structure preservation metrics and demonstrates superior or similar results on a range of clustering metrics. In terms of biological significance, {\MuST} demonstrates more accurate spot clustering, spot visualization, and spatial trajectory inference.  In addition, the gene markers found by {\MuST} are also validated by the published papers~\cite{mamoor2020alpha1, zhang2019association, maynard2021transcriptome}.

\section{Results}

\subsection{Overview of {\MuST}}

We observe that ST data suffer from the modality bias phenomenon. As shown in Fig.~\ref{fig_method_overview}A-C, there is a significant inconsistent contribution to the data label among the morphology~(Mor.) modality, transcriptomics~(Tra.) modality and spatial~(Spa.) modality. We use the interpretable method~(Shapley Value~\cite{roth1988shapley,winter2002shapley}) to show the differences in the contribution of data from different modalities to the ground truth~(Sec.~\ref{method_contributions_sharply}). As shown in Fig.~\ref{fig_method_overview}B, in the coronal mouse brain dataset (Coronal), the contribution of the Mor. modality is not consistent, and in most regions the contribution of Mor. modality is negligible. However, in some specific tissue structures (e.g., {hippocampus}), the Mor. modality provides a stronger contribution. Furthermore, the violin plots in Fig.~\ref{fig_method_overview}C show the statistical evidence of the modality bias phenomenon and it's mitigation by MuST. The modality bias phenomenon causes some methods to ignore weaker modalities such as Mor. and Spa., resulting in a loss of information. Further evidence on Fig.~\ref{fig_Mouse_Brain} and Fig.~\ref{fig_Results_Mouse_Embryos} shows that modality bias phenomenon in contributions is common in ST data~(Fig.~\ref{fig_method_overview}C). 

We propose MuST, a robust augmentation-based manifold learning ST method for several downstream tasks by fusing multiple modalities. MuST mitigates the modality bias phenomenon by introducing topology knowledge and a topology fusion loss function~(evidence in Fig.~\ref{fig_method_overview}B). MuST plays a crucial role in balancing the contributions of different modalities, thereby reducing information loss. The framework of MuST is shown in Fig.~\ref{fig_method_overview}D. Specifically, MuST first fuses the Spa. information into the Mor./Tra. spot embedding space by the Mor./Tra. encoder. Then we fuse both Mor./Tra. spot embedding spaces into the universal latent space~(fused spot embedding space) with the fusion encoder. We use a topology discovery strategy to accurately estimate global and local information for different modalities to discover the local connection in Mor./Tra. embedding spaces based on $k$-nearest neighbour ($k$NN). The superposition of local connections is then used to describe the global relationships of the data, i.e. a non-neighbouring point can be described by passing through several levels of neighbourhood relationships. To make the data distribution in the representation space more realistic, we dynamically generate new samples using neighbourhood-based interpolation and use the generated spots in the training process. Finally, we design topology fusion loss functions to fuse local and global information from different modalities using augmented data.~(Fig.~\ref{method_framework_must}).

The universal latent space learned by MuST combines local and global information from different modalities to perform downstream tasks~(Fig.~\ref{fig_method_overview}E). Specifically, a spatial clustering task using clustering methods (e.g., mclust~\cite{scrucca2016mclust}), a spot visualisation task using visualisation methods~(e.g., an MLP network based on topology loss~\cite{zang2022evnet} and UMAP~\cite{becht2019dimensionality}), marker gene analysis task using sensitivity analysis~\cite{saltelli2002sensitivity}, spatial trajectory inference task using PAGA~\cite{wolf2019paga} method, and spatial deconvolution task using Lasso~\cite{zou2006adaptive} method on the clustering results. The downstream tasks corroborate each other and provide a better understanding of the ST data. A comparison is made with several methods, including GraphST~\cite{long2023spatially}, STAGATE~\cite{dong2022deciphering}, DeepST~\cite{xu2022deepst}, SpaGCN~\cite{hu2021spagcn}, MUSE~\cite{bao2022integrative}, and stSME~\cite{pham2020stlearn} based on mean relative rank error (MRRE) score, adjusted rand index (ARI) score, and biological significance analysis.

\subsection{{\MuST} Mitigates Modality Bias and Recognizes Complex Tissues on Mouse Brain Data} \label{sec_mouse_brain}

\begin{figure}[ht]
    \centering
    \includegraphics[width=0.9\textwidth]{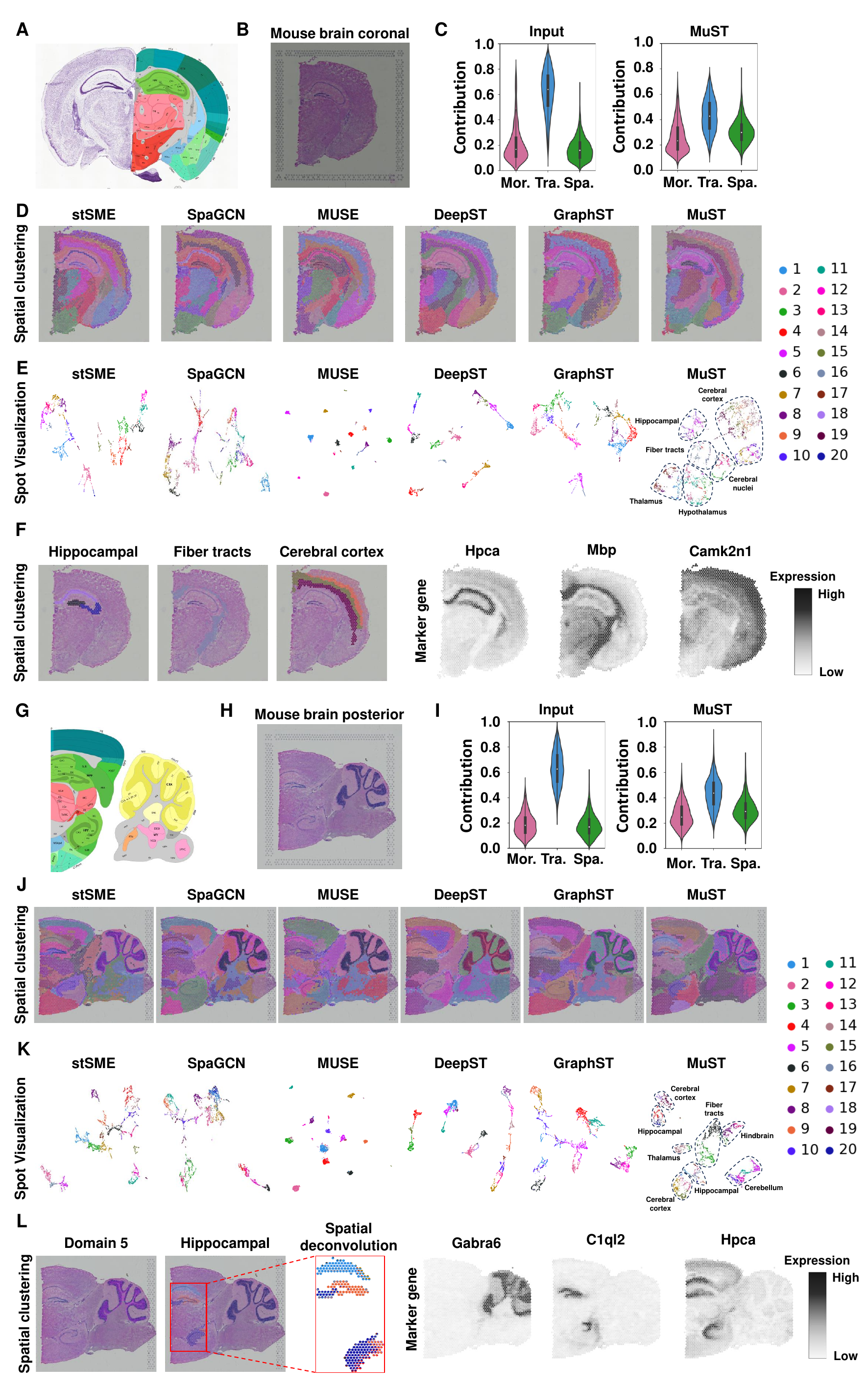}
    \caption{\textbf{{\MuST} explores more biologically complex tissues in adult mouse brain section profiled by 10x Visium (\href{https://www.10xgenomics.com/resources/datasets}{10x Genomics}). A} The annotation of hippocampus structures from the Allen Reference Atlas of an adult mouse brain. \textbf{B} H\&E image of mouse brain coronal section. \textbf{C} Contribution of the morphology (Mor.) modality, transcriptome (Tra.) modality and spatial (Spa.) modality input data or MuST embedding to data labeling. \textbf{D} Clustering results by spatial methods, stSME, SpaGCN, MUSE, DeepST, GraphST, and {\MuST}. \textbf{E} Spot visualization generated by the spatial methods. \textbf{F} Single cluster visualization of spatial domains identified by {\MuST} and the corresponding marker gene expressions. \textbf{G} Allen Brain Institute reference atlas diagram of the mouse sagittal. \textbf{H} H\&E image of mouse sagittal posterior brain section. \textbf{I} Contribution of the Mor. modality, Tra. modality and Spa. modality input data or MuST embedding to data labeling. \textbf{J} Clustering results by spatial methods, stSME, SpaGCN, MUSE, DeepST, GraphST, and {\MuST}. \textbf{K} Spot visualization generated by the spatial methods. \textbf{L} Single cluster visualization of spatial domains identified by {\MuST} and the corresponding marker gene expressions and cluster-related spatial deconvolution.}
    \label{fig_Mouse_Brain}
\end{figure}

We initially assess {\MuST}'s performance by its effectiveness in mitigating modality bias phenomenon and biological significance in mouse brain data. This evaluation confirms {\MuST}'s proficiency in accurately identifying complex tissue structures. The datasets comprise three distinct modalities: Tra., Mor., and Spa., from two specific sections of the mouse brain - coronal mouse brain section and a mouse sagittal posterior brain section (Fig.~\ref{fig_Mouse_Brain}B, H), acquired using the 10x Visium platform~\cite{10xgeomics}. We compare the spatial domains identified by {\MuST} with the corresponding anatomical reference annotations from the Allen Mouse Brain Atlas~\cite{sunkin2012allen} (Fig.~\ref{fig_Mouse_Brain}A, G), the goal is to verify our findings and assess the interpretative precision of {\MuST}. Moreover, to ensure a higher resolution in our spatial clustering task, we set the number of clusters to $20$ for both datasets. 

For the coronal mouse brain section, {\MuST} better examine the hippocampal region, fiber tract region, cerebral cortex region, and V3 structure. The spatial domains identified by baseline methods roughly partition the tissue structures into groups containing different cell types while unable to identify small spatial domains (Fig.~\ref{fig_Mouse_Brain}D). Specifically, all baseline methods fail to identify the V3 structure and ``cordlike" structure (Ammon's horn). stSME, SpaGCN, and DeepST fail to identify the ``arrow-like" structure (dentate gyrus within the hippocampus). stSME, SpaGCN, MUSE, and DeepST fail to identify some small fiber tracts. In contrast, {\MuST} brings noticeable enhancements in spatial domain identification (Fig.~\ref{fig_Mouse_Brain}D). As follows, (1) In the hippocampal region, {\MuST} identifies the CA1 field (domain 18) and CA3 field (domain 20) of the Ammon's horn and the dentate gyrus structure (domain 6). The identified hippocampal region has high concordance with the selected marker genes Hpca (Fig.~\ref{fig_Mouse_Brain}F, Tab.~\ref{tab:gene_selection_cmbs} and Fig.~\ref{fig_Results_brain_1_gene}). These advantages may be attributed to the significant contribution of Tra. modality after {\MuST} to Ammon's horn and the dentate gyrus structure (Fig.~\ref{fig_Results_Mouse_Brain_modality_bias}C). (2) {\MuST} better depicts the fiber tracts and groups the fiber tracts with texture structure in the histological image into one cluster (domain 16), which has high concordance with the selected marker genes Mbp (Fig.~\ref{fig_Mouse_Brain}F). (3) In the cerebral cortex region, {\MuST} identifies more cortical layers, including external (layers 2/3), internal (layer 4 and layer 5) and plexiform (layer 6) layers, and the thickness of the cerebral cortex is highly consistent with the anatomical reference annotations. This advantage is due to the combined contribution of Tra. modality (e.g., layers 4 and layer 6) and Mor. modality (e.g., layers 2/3 and layer 5) after {\MuST} to the cerebral cortex (Fig.~\ref{fig_Results_Mouse_Brain_modality_bias}C), thus validating the ability and significance of {\MuST} to mitigate modality bias phenomenon (Fig.~\ref{fig_Mouse_Brain}C). The identified cerebral cortex region concurs highly with the selected marker gene Camk2n1 (Fig.~\ref{fig_Mouse_Brain}F). (4) {\MuST} accurately identifies the V3 structure (domain 4), which is not possible with baseline methods. The Fabp7 gene expression is well alignment with the identified V3 structure (Fig.~\ref{fig_Results_brain_1_gene}). In the visualization results (Fig.~\ref{fig_Mouse_Brain}E), {\MuST} also separates these tissue structures marked with a dashed circle. In the spatial deconvolution task, the cluster compositions inferred by {\MuST} also accurately depict the regions identified by {\MuST} (Fig.~\ref{fig_Results_Mouse_Brain_Deconvolution}C, E).

For the mouse sagittal posterior brain section, {\MuST} better examines the hippocampal region and coronal structure. The baseline methods cannot also identify small spatial domains (Fig.~\ref{fig_Mouse_Brain}J). Specifically, they fail to identify the ``cordlike" structure, ``arrow-like" structure, and the thin layer around the coronal structure. In contrast, {\MuST} better alleviates the problems existing in these methods (Fig.~\ref{fig_Mouse_Brain}J). (1) In the hippocampal region, {\MuST} identifies the CA1 field (domain 1) and CA3 field (domain 20) of the Ammon's horn, which has high concordance with marker gene Hpca (Fig.~\ref{fig_Mouse_Brain}L, Tab.~\ref{tab:gene_selection_mpbs} and Fig.~\ref{fig_Results_brain_2_gene}). {\MuST} can group the dentate gyrus structure (domain 9) into one cluster with high concordance with the selected marker gene C1ql2 (Fig.~\ref{fig_Mouse_Brain}L). The cluster compositions inferred by {\MuST} also accurately depict the identified hippocampal region (Fig.~\ref{fig_Mouse_Brain}L and Fig.~\ref{fig_Results_Mouse_Brain_Deconvolution}D, F). These advantages may be attributed to the significant contribution of Tra. modality after {\MuST} to ``cordlike" structure and ``arrow-like" structure (Fig.~\ref{fig_Results_Mouse_Brain_modality_bias}C). (2) {\MuST} can depict the coronal structure and thin layer around it (domains 5, 11, and 12), and {\MuST} is the only method that can clearly distinguish these three domains, which correspond to the texture structure of histological image. This advantage is due to the combined contribution of Tra. modality (e.g., thin layer around the coronal structure) and Mor. modality (e.g., coronal structure) after {\MuST} to this region (Fig.~\ref{fig_Results_Mouse_Brain_modality_bias}C), thus validating the ability and significance of {\MuST} to mitigate modality bias phenomenon (Fig.~\ref{fig_Mouse_Brain}I). The Gabra6 expression aligns well with the identified spatial domain (Fig.~\ref{fig_Mouse_Brain}L). In the visualization results, {\MuST} significantly improves the separation of tissue structures (e.g., cerebellum, thalamus, cerebral cortex, hindbrain, and fiber tracts) (Fig.~\ref{fig_Mouse_Brain}K) mentioned above, this verifies that the universal latent space, which can better fuse information from multiple modalities, is beneficial for various downstream tasks.

\subsection{{\MuST} Distinctly Recognizes Fine-grained Tissue Structures on Mouse Embryo Data} \label{sec_res_embryo}

\begin{figure}[ht]
    \centering
    \includegraphics[width=0.9\textwidth]{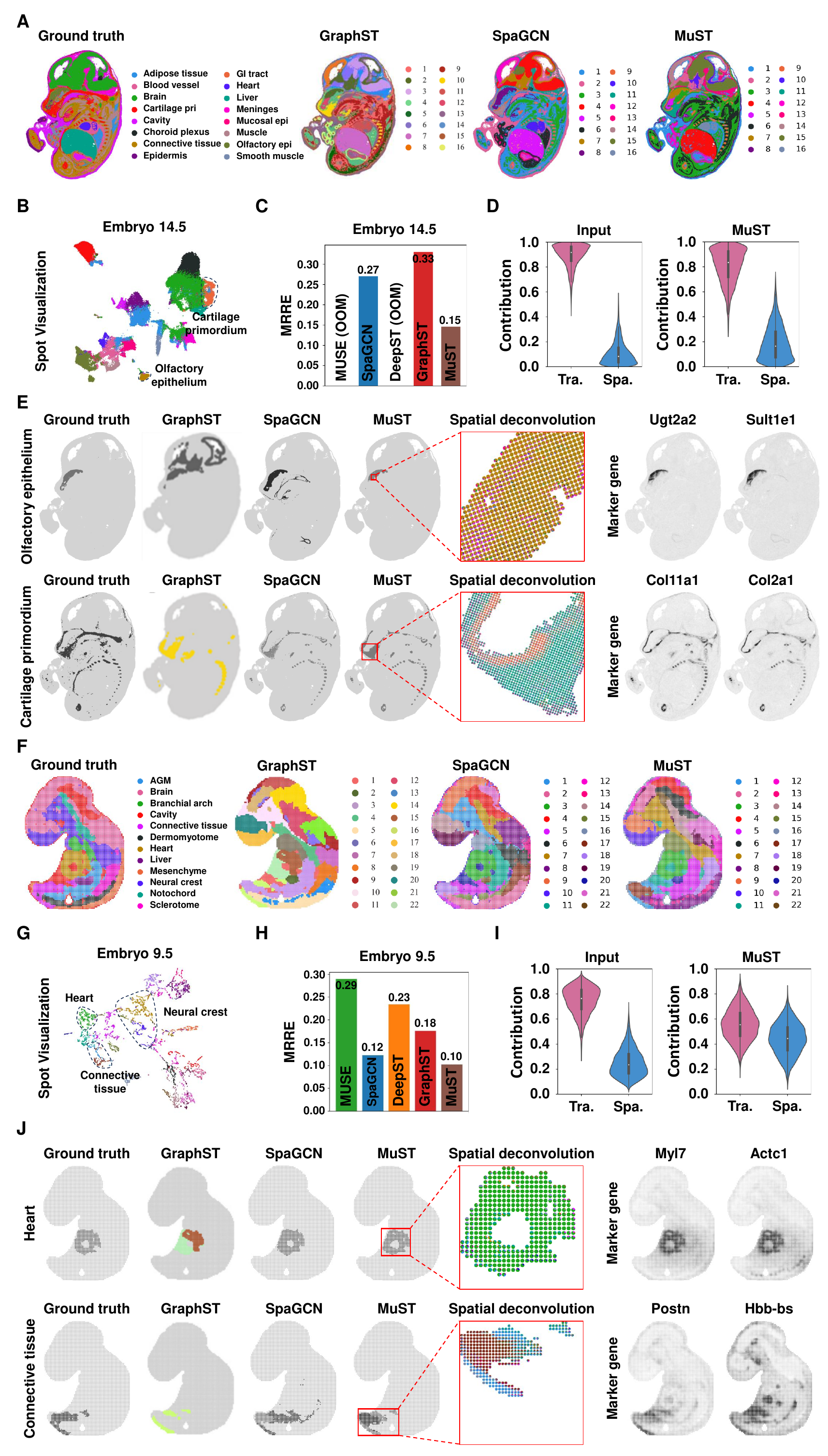}
    \caption{\textbf{{\MuST} enables accurate identification of different organs in the Stereo-seq mouse embryo (\href{https://db.cngb.org/stomics/mosta/}{MOSTA}). A} Tissue domain annotations obtained from the original Stereo-seq study and clustering results by GraphST and {\MuST} on the E14.5 mouse embryo data. \textbf{B} Spot visualizations generated by {\MuST} representations. \textbf{C} MRRE scores generated by MUSE, SpaGCN, DeepST, GraphST, and {\MuST} representations. Among them, methods MUSE and DeepST have an out-of-memory (OOM) problem. \textbf{D} Contribution of the transcriptome (Tra.) modality and spatial (Spa.) modality input data or MuST embedding to data labeling. \textbf{E} Single cluster visualization of selected spatial domains identified by the original Stereo-seq study, GraphST and {\MuST}, and the corresponding spatial deconvolution and marker genes of selected spatial domains identified by {\MuST}. \textbf{F} Tissue domain annotations obtained from the original Stereo-seq study and clustering results by GraphST and {\MuST} on the E9.5 mouse embryo data. \textbf{G} Spot visualizations generated by {\MuST} representations. \textbf{H} MRRE scores generated by MUSE, SpaGCN, DeepST, GraphST, and {\MuST} representations. \textbf{I} Contribution of the Tra. modality and Spa. modality input data or MuST embedding to data labeling. \textbf{J} Single cluster visualization of selected spatial domains identified by the original Stereo-seq study, GraphST and {\MuST}, and the corresponding spatial deconvolution and marker genes of selected spatial domains identified by {\MuST}.}
    \label{fig_Results_Mouse_Embryos}
\end{figure}

Building on our preliminary evaluation of {\MuST} in mouse brain data, which involves assessing its capability to mitigate modality bias across three modalities and its biological significance. We now broaden our research scope to encompass Stereo-seq~\cite{chen2022spatiotemporal} datasets from mouse embryos at developmental stages E14.5 (Fig.~\ref{fig_Results_Mouse_Embryos}A) and E9.5 (Fig.~\ref{fig_Results_Mouse_Embryos}F). This phase of evaluation specifically concentrates on mitigating modality bias, particularly in Tra. modality and Spa. modality, within the context of large-scale data. The tissue domain annotations for these datasets are obtained from the original study~\cite{chen2022spatiotemporal}. 

For the E14.5 mouse embryo, {\MuST} better recovers the olfactory epithelium and cartilage primordium region. We set the number of clusters to $16$ to match the original annotation for the spatial clustering task (Fig.~\ref{fig_Results_Mouse_Embryos}A). {\MuST}, GraphST and SpaGCN capture much of the fine-grained structure in the embryo and accurately identify major areas, such as the heart, liver, muscle, epidermis, and meninges regions (Fig.~\ref{fig_Results_Mouse_Embryos}A). Furthermore, {\MuST} captures more of the fine-grained structures (Fig.~\ref{fig_Results_Mouse_Embryos_145}C), such as the olfactory epithelium and cartilage primordium regions (Fig.~\ref{fig_Results_Mouse_Embryos}E). As follows, (1) {\MuST} better demarcates olfactory epithelium as a separate region~\cite{long2023spatially}, consistent with the original annotation, whereas GraphST clusters it as part of the brain region and SpaGCN assigns part of the mucosal epithelium region to it. In the visualization space (Fig.~\ref{fig_Results_Mouse_Embryos}B and Fig.~\ref{fig_Results_other_embeddings}A), {\MuST} also separates the olfactory epithelium region labeled by a dashed circle~\cite{zang2022evnet}. The identified olfactory epithelium region has high concordance with the selected marker genes Ugt2a2 and Sult1e1 (Fig.~\ref{fig_Results_Mouse_Embryos}E and Tab.~\ref{tab:gene_selection_ME145}). The cluster compositions inferred by {\MuST} accurately depict the identified olfactory epithelium region (Fig.~\ref{fig_Results_Mouse_Embryos}E). These advantages may be attributed to the significant contribution of Tra. modality after {\MuST} to olfactory epithelium region (Fig.~\ref{fig_Results_Mouse_Embryo_modality_bias}B). (2) {\MuST} better demarcates the cartilage primordium region around the brain, which has high concordance with the selected marker genes Col11a1 and Col2a1 (Fig.~\ref{fig_Results_Mouse_Embryos}E). {\MuST} also separates the cartilage primordium region labeled by a dashed circle (Fig.~\ref{fig_Results_Mouse_Embryos}B). The cluster compositions inferred by {\MuST} accurately depict the identified cartilage primordium region (Fig.~\ref{fig_Results_Mouse_Embryos}E). In addition, {\MuST} obtains the best MRRE score $0.15$ compared with GraphST ($0.33$) and SpaGCN ($0.27$) (Fig.~\ref{fig_Results_Mouse_Embryos}C).

For E9.5 mouse embryo, {\MuST} better recovers the heart, neural crest, connective tissue, and notochord regions. Although the original annotation has $12$ reference clusters (Fig.~\ref{fig_Results_Mouse_Embryos}F), the number of clusters in our testing is set to $22$ for the spatial clustering task to acquire a higher resolution of tissue segmentation~\cite{long2023spatially}. Some clusters of GraphST and SpaGCN match the annotated regions (Fig.~\ref{fig_Results_Mouse_Embryos}F), such as liver region, but the contours of mesenchyme, sclerotome, and AGM regions are not accurate. In contrast, {\MuST} further captures much of the fine-grained and contours structures, including heart, neural crest, notochord, connective tissue, brain, and cavity regions (Fig.~\ref{fig_Results_Mouse_Embryos}F and Fig.~\ref{fig_Results_Mouse_Embryos_95}C), and are highly concordant with the original annotation. As follows, (1) {\MuST} demarcates the heart region as a separate cluster, consistent with the original annotation, whereas GraphST divides it into two clusters. In the visualization results, {\MuST} also separates the heart region labeled by a dashed circle (Fig.~\ref{fig_Results_Mouse_Embryos}G and Fig.~\ref{fig_Results_other_embeddings}B). The identified heart region has high concordance with the selected marker genes Myl7, Actc1, and Acta1 (Fig.~\ref{fig_Results_Mouse_Embryos}J and Tab.~\ref{tab:gene_selection_ME95}). The cluster compositions inferred by {\MuST} accurately depict the identified heart region (Fig.~\ref{fig_Results_Mouse_Embryos}J). These advantages may be attributed to the significant contribution of Tra. modality after {\MuST} to heart region (Fig.~\ref{fig_Results_Mouse_Embryo_modality_bias}B). (2) {\MuST} better describes the contours of the notochord region and more completely demarcates the connective tissue region, whereas SpaGCN assigns part of the AGM region to connective tissue region. The identified connective tissue region concurs highly with the selected marker genes Postn and Hbb-bs (Fig.~\ref{fig_Results_Mouse_Embryos}J). This advantage is due to the combined contribution of Tra. modality and Loc. modality after {\MuST} to connective tissue region (Fig.~\ref{fig_Results_Mouse_Embryos}I and Fig.~\ref{fig_Results_Mouse_Embryo_modality_bias}B). (3) {\MuST} better demarcates the neural crest region, whereas GraphST assigns part of the mesenchyme and brain regions to this region. In addition, {\MuST} obtains best MRRE score $0.10$, followed by SpaGCN ($0.12$) and GraphST ($0.18$) (Fig.~\ref{fig_Results_Mouse_Embryos}H). The  evaluation of universal latent space representations helps us to explain the above analysis and verifies the effectiveness of modality fusing ability in {\MuST}.

\subsection{{\MuST} Distinctly Recognizes Anatomical Regions on Mouse Hippocampus Data}

\begin{figure}[ht]
    \centering
    \includegraphics[width=0.9\textwidth]{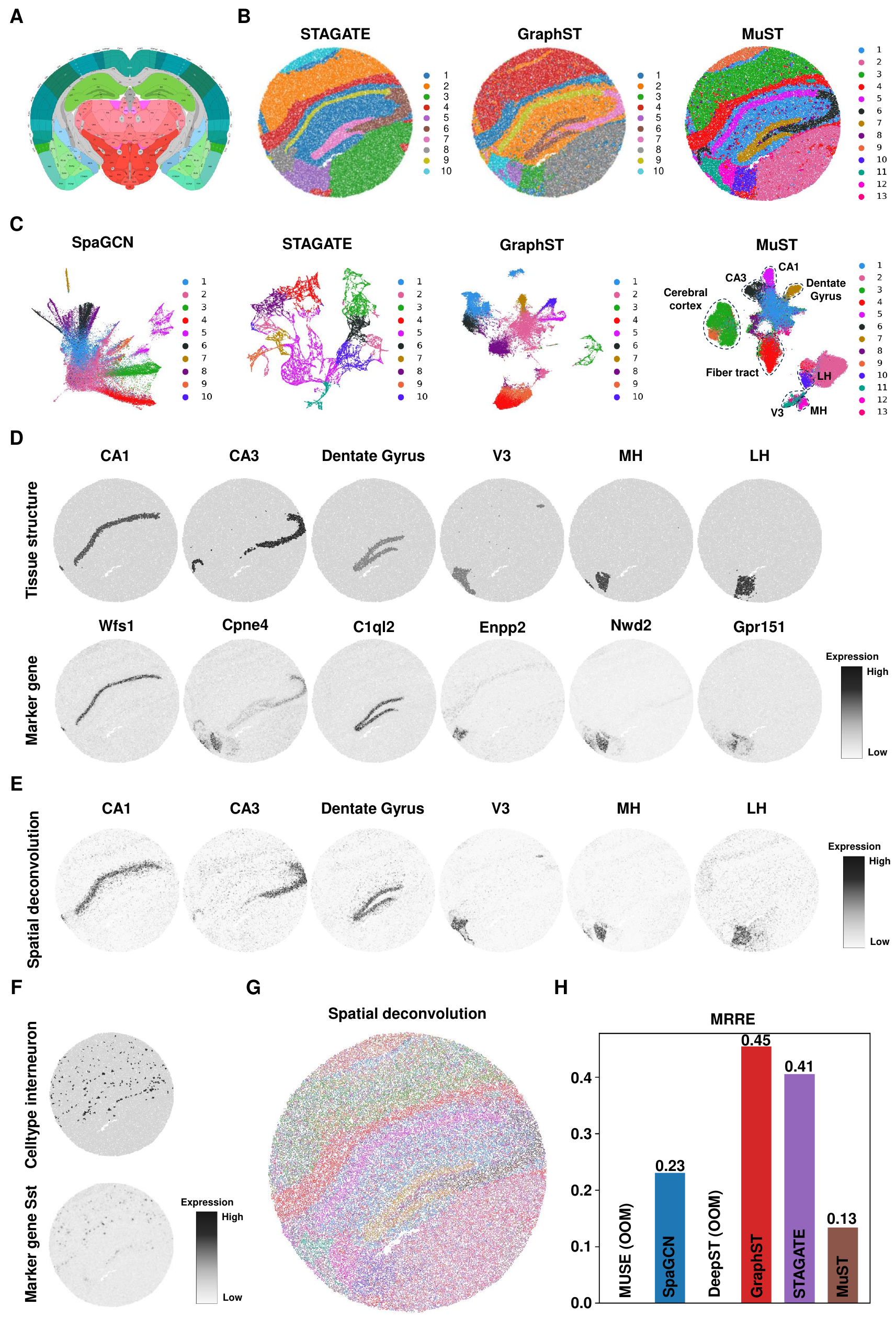}
    \caption{\textbf{{\MuST} discerns relevant anatomical regions more accurately in the SlideseqV2 mouse hippocampus data (\href{https://portals.broadinstitute.org/single_cell/study/slide-seq-study}{SCP354}). A} Allen Brain Institute reference atlas diagram of the mouse cortex. \textbf{B} Clustering results by spatial methods, STAGATE, GraphST and {\MuST} on the mouse hippocampus data. \textbf{C} Spot visualizations generated by SpaGCN, STAGATE, GraphST and {\MuST} representations. \textbf{D} Single cluster visualization of the tissue structures identified by {\MuST} and the corresponding marker gene expressions. \textbf{E} Cluster-related spatial deconvolution analysis of the tissue structures identified by {\MuST}. \textbf{F} Visualization of the celltype identified by {\MuST} and the corresponding marker gene expressions. \textbf{G} Cluster-related spatial deconvolution analysis. \textbf{H} MRRE scores generated by SpaGCN, GraphST, STAGATE, and {\MuST} representations.}
    \label{fig_Results_Mouse_Hippocampus}
\end{figure}

Building on the insights gained from the analysis of the 10x Visium and Stereo-seq platform, we extend our investigation to more advanced platform SlideseqV2~\cite{stickels2021highly}. This progression allows us to explore the consistency of spatial transcriptomic patterns across different technological iterations and the robustness of {\MuST}.
For this comparison, we employ the annotated Allen Brain Atlas as the ground truth~\cite{sunkin2012allen} (Fig.~\ref{fig_Results_Mouse_Hippocampus}A).

{\MuST} produces more spatially consistent clustering and captures major anatomical regions, such as the dentate gyrus and the pyramidal layers within Ammon's horn (CA1 and CA3 regions), and {\MuST} better discerns the cavity tissue structure and the region covered by interneuron cells. We set the number of clusters in our testing to $13$ to acquire a higher resolution of tissue segmentation (Fig.~\ref{fig_Results_Mouse_Hippocampus}B). Specifically, (1) {\MuST} and GraphST are better than STAGATE in delineating the CA3 and dentate gyrus regions with sharper boundaries. {\MuST} and STAGATE demarcate more neocortical layers in the cerebral cortex, and GraphST groups them into one region. (2) {\MuST} and GraphST can differentiate between the third ventricle (V3), medial habenula (MH), and lateral habenula (LH), which have high concordance with their marker genes Enpp2, Nwd2, and Gpr151, respectively (Fig.~\ref{fig_Results_Mouse_Hippocampus}D, Tab.~\ref{tab:gene_selection_hip} and Fig.~\ref{fig_Results_HIP_gene}). The cluster compositions inferred by {\MuST} accurately depict the identified V3, MH, and LH regions (Fig.~\ref{fig_Results_Mouse_Hippocampus}G and Fig.~\ref{fig_Results_Mouse_Hippocampus}E). However, STAGATE merges the MH and LH into one region. For the V3 structure, the Enpp2 expression does not align well with the detected V3 region of {\MuST} (Fig.~\ref{fig_Results_Mouse_Hippocampus}D), but the latter better resembles the V3 region in the annotated brain reference (Fig.~\ref{fig_Results_Mouse_Hippocampus}A). For comparison, the V3 regions of STAGATE are closer to the shape of the Enpp2 expression region but do not match the anatomical shape well. In the visualization results, {\MuST} also separates the V3, MH, and LH regions (Fig.~\ref{fig_Results_Mouse_Hippocampus}C). (3) {\MuST} can discern the cavity tissue structure and interneuron cell regions. The cavity region has higher concordance with the anatomical annotation (Fig.~\ref{fig_Results_Mouse_Hippocampus}A), and the interneuron cell region concurs more with the selected marker gene Sst (Fig.~\ref{fig_Results_Mouse_Hippocampus}F). Then, we use the MRRE score to explain the above analysis quantitatively (Fig.~\ref{fig_Results_Mouse_Hippocampus}H). {\MuST} obtains the best MRRE score $0.13$, while GraphST ($0.45$) and STAGATE ($0.41$) get much worse performance.

\subsection{{\MuST} Recognizes Laminar Structure on Mouse Olfactory Bulb Data}
Building on the discussion of the block structure analysis performed on three different platforms, we now turn our attention to evaluating the capabilities in analyzing hierarchical/laminar structures. Because a clear delineation of the laminar structure can only be achieved by reasonably fusing the information provided by different modalities.

\begin{figure}[ht]
    \centering
    \includegraphics[width=0.9\textwidth]{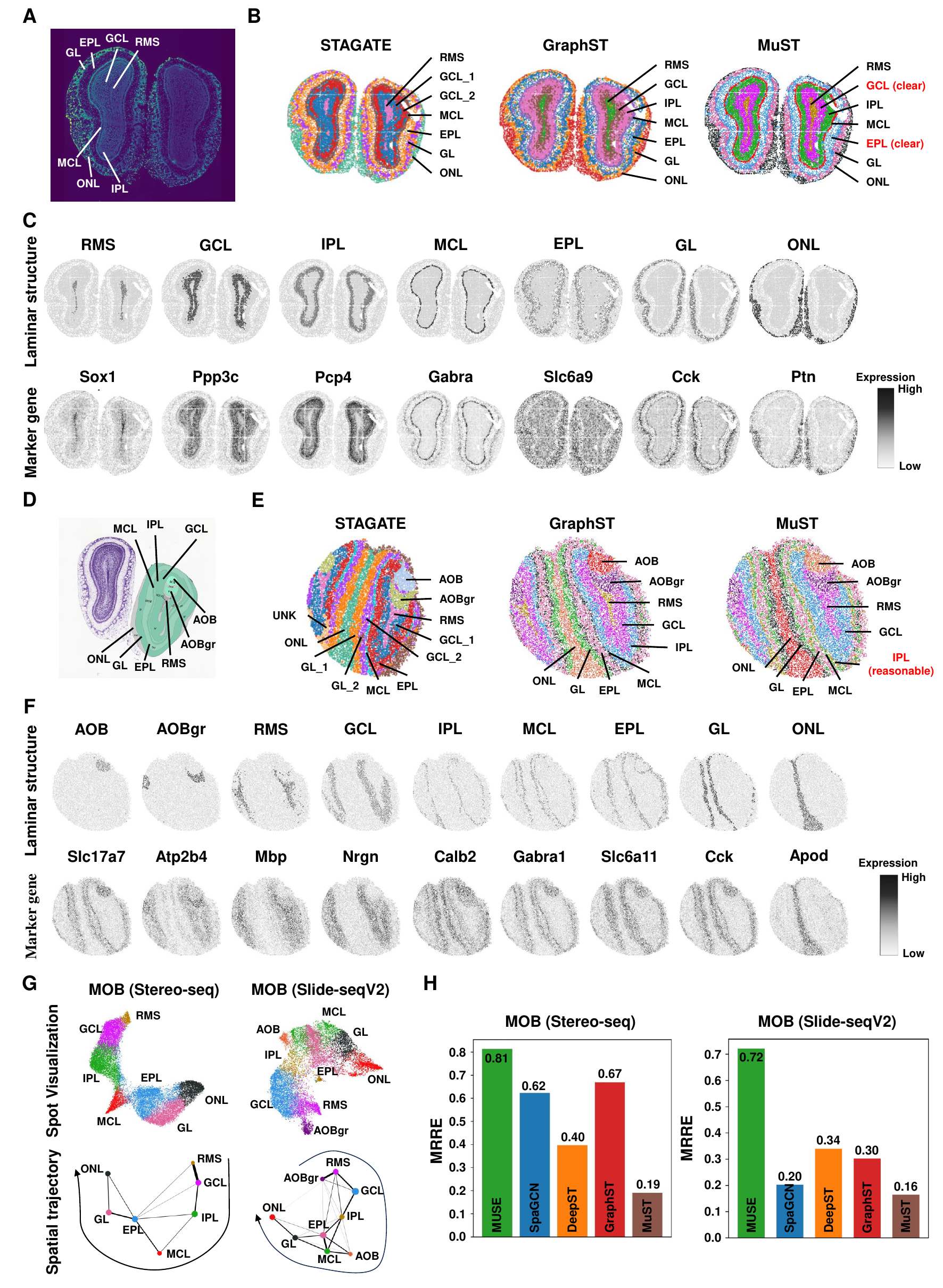}
    \caption{\textbf{{\MuST}  identifies the laminar organization in the mouse olfactory bulb tissue sections profiled by Stereo-seq (\href{https://github.com/JinmiaoChenLab/SEDR_analyses}{SEDR}) and Slide-seqV2 (\href{https://singlecell.broadinstitute.org/single_cell/study/SCP815/highly-sensitive-spatial-transcriptomics-at-near-cellular-resolution-with-slide-seqv2}{SCP815}) respectively. A} Laminar organization of mouse olfactory bulb annotated in the DAPI-stained image generated by Stereo-seq. \textbf{B} Clustering results by spatial methods, STAGATE, GraphST and {\MuST} on the Stereo-seq mouse olfactory bulb tissue section. \textbf{C} Single cluster visualization of the spatial domains identified by {\MuST} and the corresponding marker gene expressions. \textbf{D} Laminar organization of mouse olfactory bulb annotated by the Allen Reference Atlas. \textbf{E} Clustering results by spatial methods, STAGATE, GraphST and {\MuST} on Slide-seqv2 mouse olfactory bulb tissue section. \textbf{F} Single cluster visualization of spatial domains identified by {\MuST} and the corresponding marker gene expressions. \textbf{G} Spot visualizations and PAGA graphs generated by the representations of {\MuST}. \textbf{H} MRRE scores generated by MUSE, SpaGCN, DeepST, GraphST, and {\MuST} representations.}
    \label{fig_Mouse_Brain_Olfactory_Bulb}
\end{figure}

We use two mouse olfactory bulb tissue datasets acquired with high-resolution platforms Stereo-seq~\cite{chen2022spatiotemporal} and Slide-seqV2~\cite{stickels2021highly}, to rigorously test {\MuST}'s proficiency in delineating intricate laminar architectures. For the former, we compare the spatial domains identified by {\MuST} with the mouse olfactory bulb's laminar structure annotated by the DAPI-stained image~\cite{fu2021unsupervised, dong2022deciphering}, identifying the olfactory nerve layer (ONL), glomerular layer (GL), external plexiform layer (EPL), mitral cell layer (MCL), internal plexiform layer (IPL), granule cell layer (GCL), and rostral migratory stream (RMS) (Fig.~\ref{fig_Mouse_Brain_Olfactory_Bulb}A). For the latter, we compare the spatial domains identified by {\MuST}  with the laminar organization of mouse olfactory bulb annotated by the Allen Reference Atlas~\cite{sunkin2012allen}, identifying the accessory olfactory bulb (AOB), granular layer of the accessory olfactory bulb (AOBgr), RMS, GCL, IPL, MCL, EPL, GL, and ONL (Fig.~\ref{fig_Mouse_Brain_Olfactory_Bulb}D).

For mouse olfactory bulb tissue section acquired with Stereo-seq, {\MuST} more clearly recognizes the external and internal structure of the organ. We set the number of clusters to $7$ to match the known laminar structure for the spatial clustering task. STAGATE, GraphST, and {\MuST} can separate the outer layers of the organ, namely the ONL, GL, and EPL, but STAGATE seriously confuses GL and EPL structures (Fig.~\ref{fig_Mouse_Brain_Olfactory_Bulb}B). {\MuST}  and GraphST can demarcate the GCL and RMS structures for the inner structure. However, GraphST's clusters are noisy, and the boundaries between different structures are not as straightforward as {\MuST}. We use the respective selected marker genes of each anatomical region to validate {\MuST}'s clusters (Fig.~\ref{fig_Mouse_Brain_Olfactory_Bulb}C), and find the excellent correspondence between {\MuST}'s clusters and the known marker genes (Tab.~\ref{tab:gene_selection_mob_stereo} and Fig.~\ref{fig_Results_MOB_Stereo_gene}), such as Gabra1~\cite{mamoor2020alpha1}, Cck and Ptn. For some marker genes (e.g., Ppp3ca and Pcp4), their high expression levels overlap with neighboring regions. This is expected as cell types are often shared among the different inner structures of organs, and markers are likewise shared among similar cell types. {\MuST} delineates the spatial trajectory among the mouse olfactory bulb (from RMS to GCL to ONL) with clear boundaries between different structures in the visualization results as well as the PAGA graphs~\cite{wolf2019paga} (Fig.~\ref{fig_Mouse_Brain_Olfactory_Bulb}G), but other methods do not have this capability (Fig.~\ref{fig_Results_other_embeddings}C). In addition, {\MuST} obtains the best MRRE score 0.19 (Fig.~\ref{fig_Mouse_Brain_Olfactory_Bulb}H), while MUSE (0.81) and GraphST (0.67) get much worse performance.

For mouse olfactory bulb tissue section acquired with Slide-seqV2, {\MuST} better recognizes the laminar structure using fewer clusters. We set the number of clusters to $9$ to match the known laminar structure for the spatial clustering task. The spatial domains identified by {\MuST} are consistent with the mouse olfactory bulb annotation from the Allen Reference Atlas. At the same time, baseline methods need more clusters to identify these structures (Fig.~\ref{fig_Mouse_Brain_Olfactory_Bulb}E). These spatial domains uncovered by {\MuST} are clearly supported by known and selected gene markers (Fig.~\ref{fig_Mouse_Brain_Olfactory_Bulb}F, Tab.~\ref{tab:gene_selection_mob_slide} and Fig.~\ref{fig_Results_MOB_Slide_gene}). The granular cell marker Atp2b4~\cite{zacharias1999developmental} shows strong expressions on the identified AOBgr domain. {\MuST} identifies the GCL structure with the dominant expression of Nrgn~\cite{zhang2019association}. Nrgn is a well-documented schizophrenia risk gene, implying that this domain is related to cognition function. The narrow MCL structure with the dominant expression of mitral cell marker Gabra1~\cite{mamoor2020alpha1} is also identified by {\MuST}. {\MuST} delineates the spatial trajectory among the mouse olfactory bulb (from AOBgr to RMS to ONL) in the visualization results and the PAGA graphs (Fig.~\ref{fig_Mouse_Brain_Olfactory_Bulb}G), but other methods do not have this capability (Fig.~\ref{fig_Results_other_embeddings}D). In addition, {\MuST} obtains the best MRRE score of 0.16, while MUSE (0.72) and GraphST (0.48) get much worse performance (Fig.~\ref{fig_Mouse_Brain_Olfactory_Bulb}H). Collectively, these results illustrate the modalities fusing ability of {\MuST} to identify tissue structures and reveal their organization from ST data of different spatial resolutions.

\subsection{{\MuST} Exploratively Recognizes Layers on Human Dorsolateral Prefrontal Cortex Data}\label{sec_DLPFC_151673}

\begin{figure}[ht]
    \centering
    \includegraphics[width=0.9\textwidth]{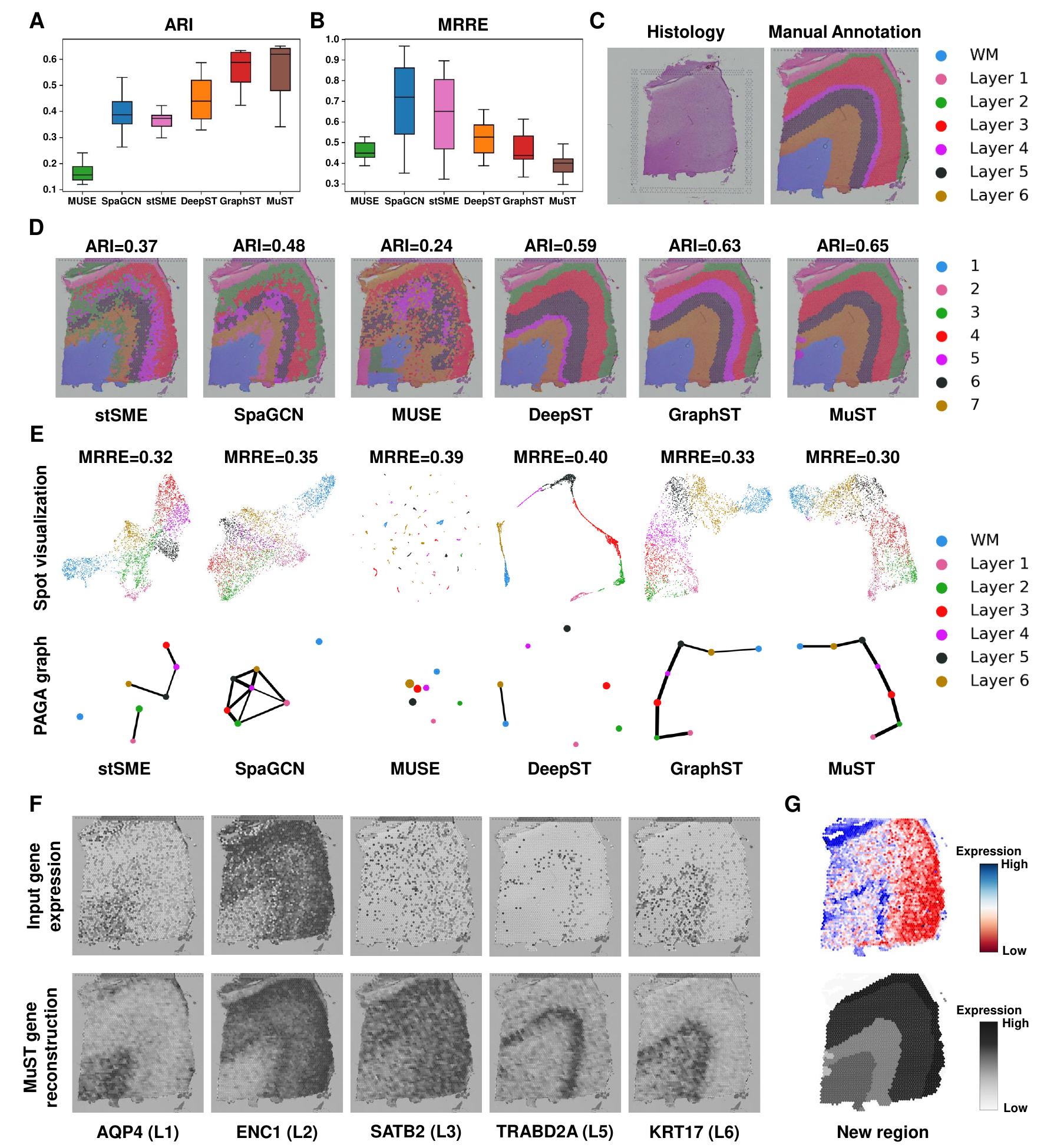}
    \caption{\textbf{{\MuST}  improves the identification of layer structures in the human dorsolateral prefrontal cortex (DLPFC) tissue (\href{http://research.libd.org/spatialLIBD/}{spatialLIBD}). A} Boxplots of adjusted rand index (ARI) scores of six spatial methods applied to the 12 DLPFC slices. Boxplots denote medians and interquartile ranges (IQRs). The whiskers of a boxplot are the lowest datum still within 1.5 IQR of the lower quartile and the highest datum within 1.5 IQR of the upper quartile. \textbf{B} Boxplots of mean relative rank error (MRRE) scores of six spatial methods applied to the 12 DLPFC slices. \textbf{C} H\&E image and manual annotation from the original study. \textbf{D} Clustering results by spatial methods, stSME, SpaGCN, MUSE, DeepST, GraphST, and {\MuST} on slice 151673 of the DLPFC dataset. Manual annotations and clustering results of the other DLPFC slices are shown in supplementary information Fig.~\ref{fig_Results_DLPFC_all}. \textbf{E} Spot visualizations and PAGA graphs generated by stSME, SpaGCN, MUSE, DeepST, GraphST, and {\MuST} representations on slice 151673. \textbf{F} Visualizations of the input gene expressions and {\MuST} gene reconstructions of five layer-marker genes on slice 151673. \textbf{G} Gene expressions analysis of the new regions.}
    \label{fig_DLPFC_151673}
\end{figure}

Having established {\MuST}'s efficacy in analyzing block structures in mouse data, we extend our investigation to human datasets to assess its broader applicability. This transition from model organisms to human data is crucial for validating {\MuST}'s utility in more complex and clinically relevant settings.

We apply {\MuST} to LIBD human dorsolateral prefrontal cortex (DLPFC) dataset~\cite{maynard2021transcriptome}, which contains ST profiles of $12$ DLPFC slices and each depicting the four or six layers of the human dorsolateral prefrontal cortex and white matter (WM). Across $12$ slices, for ARI score (Fig.~\ref{fig_DLPFC_151673}A), {\MuST} achieves the highest median score $0.62$, GraphST obtains the second median score $0.59$ after {\MuST}, MUSE obtains the poorest median score $0.16$, and the remaining methods have median ARI scores of less than $0.44$. For MRRE score (Fig.~\ref{fig_DLPFC_151673}B), {\MuST} achieves best median score $0.40$, GraphST obtains the second median score $0.44$, and the MRRE median scores of SpaGCN ($0.72$) and stSME ($0.65$) are worse. We illustrate the results with one slice 151673 (Fig.~\ref{fig_DLPFC_151673}D-G).

{\MuST} achieves more competitive performance on spatial clustering tasks than baseline methods. The visual comparisons clearly show that the MUSE performs the worst spatial clustering, failing to identify every slice layer 151673. stSME and SpaGCN slightly alleviate the problem existing in MUSE with the ability to recover WM, but with the remaining clusters mixed among six layers, including the WM has a broken boundary and being clustered together with a portion of Layer 6, and the boundaries between other clusters are also very chaotic with no clean separation. DeepST accomplishes better separation of each layer, but with incorrect layer thickness, fails to recover Layer 4, and the boundaries between WM and Layer 6 still need to be more apparent. GraphST and {\MuST} produce layers that are significantly closer in shape to the manual annotation (Fig.~\ref{fig_DLPFC_151673}C), but with some minor flaws. GraphST fails to recover Layer 2, and the thickness of Layer 1 and Layer 4 are not accurate, while {\MuST} successes to recover Layer 2 and better captures the position and thickness (Fig.~\ref{fig_DLPFC_151673}D). {\MuST} fails to identify Layer 4, which is understandable because Layer 4 is usually skinny and contains similar cell types to Layer 3, such as pyramidal neurons. Therefore, {\MuST} discovers a new spatial domain that differs significantly in gene expression from the others (Fig.~\ref{fig_DLPFC_151673}G). For quantitative assessment, {\MuST} achieves the highest ARI score of 0.65, GraphST obtains the second highest ARI score of 0.63, while the MUSE is the poorest method at an ARI score of 0.24. The results with all other slices are shown in Fig.~\ref{fig_Results_DLPFC_all}.

{\MuST} effectively incorporates Spa. modality in the data, which allows it to reflect the distance between spatial domains and capture spatial trajectory in the visualization results~\cite{becht2019dimensionality}. For example, in slice 151673, the different cortical layers appear well-organized and follow a clear spatial trajectory (from Layer 1 to Layer 6 to WM) in the visualization results (Fig.~\ref{fig_DLPFC_151673}E). This result is consistent with the functional similarity between adjacent cortical layers and the chronological order~\cite{gilmore1997cortical}. By contrast, in the visualization results of MUSE embeddings, spots belonging to the same layers are split into multiple clusters. As for the other spatial clustering methods, stSME mixes the spots of layer 1 (green) and other cortex layers, and SpaGCN does not distinguish Layer 4 and other cortex layers clearly. We additionally validate the inferred spatial trajectory utilizing PAGA (Fig.~\ref{fig_DLPFC_151673}E). The PAGA graphs of {\MuST} and GraphST representations show an almost linear development trajectory from Layer 1 to Layer 6 and the similarity between adjacent layers. In contrast, the PAGA result of both SpaGCN representations is mixed to different degrees, and the PAGA results of stSME, MUSE, and DeepST representations cannot recover spatial trajectories. For quantitative assessment, {\MuST} achieves the lowest MRRE score of 0.30, while the DeepST is the poorest method at an MRRE score of 0.40. In addition, {\MuST} can denoise the gene expressions based on the universal latent space. We apply {\MuST} to decrease noises in the DLPFC dataset to better show the spatial pattern of genes across the layers of the cortex. We compare the expression profiles of five layer-marker genes of the input data to the denoised representations by {\MuST} in the slice 151673 in Fig.~\ref{fig_DLPFC_151673}F. As anticipated, the {\MuST}-denoised representations clearly show the laminar enrichment of these layer-specific marker genes. For example, after denoising, the AQP4, ENC1, SATB2, TRABD2A, and KRT17 genes~\cite{maynard2021transcriptome} show differential expressions in Layer 1, Layer 2, Layer 3, Layer 5, and Layer 6 respectively, which are consistent with previously reported results.

\section{Discussion}
This paper introduces {\MuST}, a augmentation-based manifold learning model that integrates gene expression, tissue morphology, and spatial information into a latent model for a variety of downstream tasks, including spatial clustering~\cite{long2023spatially, bao2022integrative}, spot visualization~\cite{dries2019giotto,xu2023structure}, spatial deconvolution~\cite{kleshchevnikov2022cell2location, ma2022spatially}, marker gene analysis~\cite{dumitrascu2021optimal}, and spatial trajectory inference~\cite{wolf2019paga}, with improved precision. 

{\MuST} mitigates the detrimental effects of modality bias and therefore has a robust modality integration capability. The information in the weaker modalities (Mor. modality and Spa. modality) is better exploited to characterise and improve the performance of various downstream tasks. As an indication of the benefits of these techniques, MuST exploits morphological and textural differences to improve the identification of key tissue structures. For example, in the coronal mouse brain dataset and the sagittal mouse posterior brain dataset (Sec.~\ref{sec_mouse_brain} and Fig.~\ref{fig_Mouse_Brain})~\cite{10xgeomics}, {\MuST} more accurately detects morphological variations, which sharpens the distinction of hippocampal region, cerebral cortex region and coronal structure.
However, GraphST~\cite{long2023spatially} fails to identify the ``string-like'' and ``arrow-like'' structures due to its neglect of morphological information. The coarse-grained detection of hippocampal structure by other methods (e.g., stSME~\cite{pham2020stlearn}, SpaGCN~\cite{hu2021spagcn} and DeepST~\cite{xu2022deepst}) may be due to the negative effect of the modality bias. Furthermore, in the human DLPFC dataset (Sec.~\ref{sec_DLPFC_151673} and Fig.~\ref{fig_DLPFC_151673})~\cite{maynard2021transcriptome}, {\MuST} more effectively recognises the texture structure (cluster 2 in Fig.~\ref{fig_DLPFC_151673}D) in histology images, while GraphST neglects this information. MuST is also better at recovering the fine-grained tissue structures in mouse embryo datasets due to the integration of Spa. and Tra. modalities (Sec.~\ref{sec_res_embryo} and Fig.~\ref{fig_Results_Mouse_Embryos}). It handles large data sets, although the stereo-seq technology does not provide data for the Mor. modality. However, the modality bias leads to inaccurate results from other methods (e.g. olfactory epithelium, cartilage primordium, heart, neural crest, notochord, connective tissue, brain and cavity regions) compared to the ground truth.

Another advantage of {\MuST} is its ability to perform synergistic analysis across multiple downstream tasks, potentially yielding consistent, confirmatory results. In spatial clustering, {\MuST} identifies distinct regions such as olfactory epithelium, cartilage primordium, heart, notochord and connective tissue in embryo data~(Sec.~\ref{sec_res_embryo}), and hippocampal areas (e.g. CA1, CA3, dentate gyrus), fibre tracts, cerebral cortex and V3 structures in mouse brain data~(Sec.~\ref{sec_mouse_brain}). At the same time, some regions (e.g. olfactory epithelium, hippocampus, fibre tracts, cerebral cortex) are also delineated in the visualisation subtask, revealing intricate details of tissue similarity. Building on this universal latent space, spatial clustering and spot visualisation guide the interpretability and feature selection task to focus on specific genes associated with the identified differences. For example, the unity between the tagged regions and selected marker genes - Ugt2a2 and Sult1e1 in the olfactory epithelium region, Myl7 and Actc1 in the heart region~(Sec.~\ref{sec_res_embryo}), and C1ql2 and Hpca in the hippocampus region~(Sec.~\ref{sec_mouse_brain}) - is evident. The cluster compositions derived from {\MuST} accurately characterise these regions. These downstream tasks show remarkable consistency in biological significance, outperforming other baseline methods.

Ultimately, {\MuST} is versatile, processing ST data from a variety of platforms, such as 10x Visium~\cite{hudson2022localization}, Slide-seqV2~\cite{stickels2021highly} and Stereo-seq~\cite{chen2022spatiotemporal}, and effectively managing the different spatial resolutions inherent in these technologies. In the future, we plan to extend {\MuST} to build a pre-trained base model to provide a foundation not only for the tasks considered in this work, but also for other related tasks, including solving batch effect problems through graph structure integration, annotating cell types, and interpreting developmental trajectories. Due to its broad applicability, flexibility and potential for extension, we anticipate that {\MuST} will become an indispensable tool in ST research.

\section{Methods}

\subsection{Data Preprocessing}

Spatial transcriptomics (ST) data include gene expression counts~(transcriptomics modality), optional tissue morphology images~(morphology modality), and spatial coordinates~(spatial modality). First, as shown in Fig.~\ref{fig_method_framework}A, gene expression counts are log transformed and normalised for library size using SCANPY~\cite{wolf2018scanpy}. These counts are then standardised to zero mean and unit variance. The next step is to select the 3000 most variable genes~\cite{svensson2018spatialde} for the {\MuST} model input. For tissue images, as shown in Fig.~\ref{fig_method_framework}B, we centre and crop a $224 \times 224$ pixel area around the probe coordinates. These images are then processed by the ResNet50 architecture~\cite{kaiming2016resnet}, pre-trained on the ImageNet collection, to distill a 2048-dimensional feature vector, excluding the final classification layer. To further refine the data, we apply PCA, capturing the top 50 principal components to represent spot morphology.

The ST data comprised two distinct components: transcriptomic~(Tra.) modality $\bm{{X}}^\text{tr}$, encapsulating the refined gene expression profiles, and morphology~(Mor.) modality $\bm{{X}}^\text{mo}$. The Tra. modality $\bm{{X}}^\text{tr}=\{\bm{{x}}^\text{tr}_n\}_{n=1}^{N}$ consists of vectors $\bm{{x}}^\text{tr}_n$ in the space $\mathbb{R}^{{N_t}}$, with $N_t$ representing the count of selected genes. The Mor. modality $\bm{{X}}^\text{mo} = \{\bm{{x}}^\text{mo}_n\}_{n=1}^{N}$ encompasses the histological image features, with $\bm{{x}}^\text{mo}_n$ belonging to $\mathbb{R}^{N_m}$ and $N_m$ denoting the feature count. To facilitate the analysis of this multiple modalities dataset, we construct a heterogeneous graph $\bm{G}(\{\bm{E}^{p}\}, \{\bm{X}^\text{tr}, \bm{X}^\text{mo}\})$, where $\bm{E}^{p}$ are the sets of edges interlinking spots across spatial dimensions. We use an $\epsilon$-radius approach~\cite{haggstrom1998nearest}, connecting points that are within a certain distance $\epsilon$.  

\subsection{Detecting modality's contributions to labels using Shapley Value}\label{method_contributions_sharply}

Modality bias phenomenon in ST data refers to the tendency of data to be overly biased towards one domain, often resulting from attempts to align features across different domains, leading to a loss of domain-specific discriminative information\cite{gat2020removing,guo2023modality}.
We use, Shapley Value\cite{roth1988shapley,winter2002shapley}, an interpretable method to determine how much the data for each modality contributes to the true label. In practice, we use the shap package\footnote{https://github.com/shap/shap} directly to explain a default linear SVC \footnote{https://scikit-learn.org/stable/modules/generated/sklearn.svm.LinearSVC.html} model that is trained using multiple modalities data~(or single modalities embedding of MuST) as input and real labels as output. We extract the resulting shap values and merge them by modality via `max()' as a contribution from each modality.

\subsection{Framework of {\MuST}}\label{method_framework_must}

To obtain a universal latent space that fuses information from different modalities to support various downstream applications, we design the {\MuST} framework~(Fig.~\ref{fig_method_framework}C). The {\MuST} framework integrates, a data augmentation module $\mathcal{A}(\cdot)$, a multi-modality fusion network $\mathcal{F}(\cdot)$, and a decoding network $\mathcal{D}(\cdot)$. In addition, a topology fusion loss $\mathcal{L}_\text{T}(\cdot)$ and a reconstruction loss function $\mathcal{L}_\text{R}(\cdot)$ are used to effectively train these components.

\textbf{Data Augmentation Module.}
The data augmentation module $\mathcal{A}(\cdot)$ produces novel augmented samples by exploiting the constructed heterogeneous graph. As shown in Fig.~\ref{fig_method_framework}D, augmented instances $\bm{x}^+$ are created to match the distribution of data across multiple modalities. The module generates new data by integrating the topological information with the input features.

For data $x \in \mathbb{R}^N$, with $N$ representing the feature count, the data augmentation module's operation is defined as:
\begin{equation}
	\begin{aligned}
		 & \bm{x}^+ =  \mathcal{A}_\mathcal{U}(\bm{x})  = (1-r_u) \cdot \bm{x} + r_u \cdot \bm{x}^{\text{hop}^1} , \\
		 & \bm{x}^{\text{hop}^1} \sim \text{Hop}^1(x, E), r_u \sim U(0, p_U)
	\end{aligned}
\end{equation}
where $\bm{x}^{\text{hop}^1}$ is drawn from the 1-hop neighbours of $\bm{x}$ within the edge set $E$. The coefficient $r_u$ is the mixing parameter, while $p_U$ is the uniform distribution hyperparameter. The augmentation operations are performed in real time during network training, increasing the variability of the data and preserving the integrity of the feature semantics. During training, $\mathcal{H}_{ij}=0$ indicates the augmentation relationship between $\bm{x}_j$ and $\bm{x}_i$. For example, if $\mathcal{H}_{ij}=1$ then $\bm{x}_j$ is the augmented data of $\bm{x}_i$, id $\mathcal{H}_{ij}=0$, $\bm{x}_j$ is sampled from the same modality data set.
\begin{equation}
	\begin{aligned}
		\bm{x}_i  \in \bm{X}^\text{m},
		\bm{x}_j  \sim
		\left\{
		\begin{aligned}
			 & \bm{x}_j  \sim \bm{X}^\text{m}       & \text{if} \ \ \  \mathcal{H}_{ij}=0 \\
			 & \bm{x}_j  \sim \mathcal{A}(\bm{x}_i) & \text{if} \ \ \  \mathcal{H}_{ij}=1
		\end{aligned}
		\right.
		\label{eq_CL}
	\end{aligned}
\end{equation}

\textbf{Multi-modality Fusion Network.}
The multi-modality fusion network includes Mor. Encoder, Tra. Encoder and a Fusion Encoder.
The Mor. Encoder, Tra. Encoders are based on Graph Neural Networks~(GNN)~\cite{kipf2017semisupervised} and extract local structural features from the Mor. and Tra. modalities using spatial coordinates. The Fusion Encoders are based on Multilayer Perceptron (MLP)~\cite{tang2015extreme} and fuse the information from the output of the Mor. Encoder and Tra. Encoder. The fusion process is described by the following equation:
\begin{equation}
	\begin{aligned}
		\bm{z}           & = \text{MLP}_{\phi}(\bm{y}), \bm{y} = \text{Enc}_{\pi}(\text{cat}(\bm{y}^\text{mo}, \bm{y}^\text{tr})) \\
		\bm{y}^\text{mo} & = \text{GNN}_{\theta}(\bm{x}^\text{mo}, E^\text{p})                                                   \\
		\bm{y}^\text{tr} & = \text{GNN}_{\omega}(\bm{x}^\text{tr}, E^\text{p})
	\end{aligned}
\end{equation}
where $\bm{x}^\text{mo}$ and $\bm{x}^\text{tr}$ represent the Mor. and Tra. data inputs, while $E^\text{p}$ denotes the spatial coordinate graph structure. The resultant vector $\bm{z}$ resides in the Fused Spot embedding space (Fused Spot Emb). The vectors $\bm{y}^\text{mo}$ and $\bm{y}^\text{tr}$ occupy the Mor. and Tra. spot embedding spaces, respectively. The operation $\text{cat}$ denotes the concatenation of data. $\text{Enc}_{\pi}$ is the dedicated encoder for data fusion. The GNNs, $\text{GNN}_{\theta}$ and $\text{GNN}_{\omega}$, process Mor. and Tra. data, respectively. 

\textbf{Multi-modality Fusion Network.}
The multi-modality fusion network is used to reconstruct the input data from the latent space:
\begin{equation}
	\begin{aligned}
		\bm{\hat{x}}^\text{tr} & = \mathcal{D} (\bm{z}) =\text{Dec}_{\psi}(\bm{z})
	\end{aligned}
\end{equation}
where $\bm{\hat{x}}^\text{tr}$ is the reconstructed Tra. data, and $\text{MLP}_{\psi}$ is the multilayer perceptron for Tra. data reconstruction.

\textbf{Topology Discovery.}\label{Graph_Construction_method}
To mitigate the adverse effects of modality bias phenomenon, MuST does not directly fuse the information of modalities, instead of measuring the importance of modality in Mor. and transcriptomic (Tra.) spot embedding space (Fig. \ref{fig_method_framework}D). Specifically, we proposed a topology discovery strategy to accurately estimate global and local information of multi-modality data by extracting the topology from Mor./Tra. spot embedding space. 
The $E^\text{tr}$ and $E^\text{mo}$ are discovered with $k$-nearest neighbour ($k$NN~\cite{peterson2009k}) according to the distance in the Mor./Tra.spot embedding space. 

\textbf{Topology Fusion Loss.} 
In addition, a topology fusion loss is proposed to minimize the topological differences between all modalities and the latent space with an adaptive motility's importance estimated from the continuously updated Mor./Tra. 's spot embedding:
\begin{equation}
	\begin{aligned}
		\mathcal{L}_T(\bm{X}, \bm{y}, \bm{z})                          & = \sum_{\text{m}\in \mathcal{M}} \mathcal{L}_T^\text{m}(\bm{X}^\text{m}, E^\text{m},\bm{y}, \bm{z}) \\
		\mathcal{L}_T^\text{m}(x^\text{m}, \bm{E}^\text{m}, \bm{y}, \bm{z}) & = \sum \{
		\mathcal{T}_{ij}^\text{m}  \log(S_{ij}) +(1-\mathcal{T}_{ij}^\text{m})\log(1-S_{ij})
		\},                                                                                                                                   \\
	\end{aligned}
\end{equation}
where $\mathcal{M}$ is the set of modalities, in this paper, $\mathcal{M} = \{\text{tr}, \text{mo}\}$. $\bm{X}^\text{m}$ and $E^\text{m}$ are the data and topological structure of modality $\text{m}$. $S_{ij}$ is used to describe the connection strength between data $\bm{z}_i$ and $\bm{z}_j$:
\begin{equation}
	\begin{split}
		{S}_{ij} &= \kappa \left({\bm{z}}_i, {\bm{z}}_j \right), \\
		\kappa(a, b) &= (1 + || a - b ||^2 / \nu)^{-\frac{\nu+1}{\nu}}
	\end{split}
\end{equation}
where the kernel function $\kappa(a,b)$ maps the relationship of the input data to the similarity, and the degree of freedom $\nu$ is related to the number of the dimension.
$\mathcal{T}_{ij}^\textbf{m}$ is the topological priory between vector $\bm{y}_i^\text{m}$ and $\bm{y}_j^\text{m}$:
\begin{equation}
	\begin{split}
		{\mathcal{T}}_{ij}^\text{m} =
		\left[1 + \mathcal{H}_{ij}^\text{m}\left(e^\alpha-1\right)\right] \kappa(\bm{y}_i^\text{m}, \bm{y}_j^\text{m})
	\end{split}
\end{equation}

\textbf{Overall Loss Function} Our overall loss function is formulated as follows:
\begin{equation}
	\mathcal{L} = \mathcal{L}_T(\bm{X}) + \lambda \mathcal{L}_R(\bm{x}^\text{tr}, \hat{\bm{x}}^\text{tr})
\end{equation}
where $\lambda$ is a hyper-parameter tuning the influence between two loss functions. 
The reconstruction loss $\mathcal{L}_R(\bm{x}^\text{tr}, \hat{\bm{x}}^\text{tr})$ is designed to ensure that the $\mathcal{F}(\cdot)$ loses as little information about the gene modality as possible:
\begin{equation}
	\mathcal{L}_R(\bm{x}^\text{tr}, \hat{\bm{x}}^\text{tr}) = \frac{1}{N} \sum^N_{i=0}(\bm{x}^\text{tr}_i -  \hat{\bm{x}}^\text{tr}_{i})^2
\end{equation}

\subsection{Downstream Tasks of {\MuST}}

The proposed {\MuST} is dedicated to learning a unified embedding space, which is used for several downstream tasks. These downstream tasks are used in combination to confirm the results of ST analyses. The predefined downstream tasks of {\MuST} are described below:

\textbf{Spatial clustering.} This task focuses on the clustering of spots based on the embedding space. It aids in identifying regions of interest or patterns that might be indicative of specific tissue structure or cell type. Spatial clustering accomplishes clustering according to a universal latent space, and various popular clustering methods can be employed. For fairness of comparison with the baseline methods, we show the clustering results of mclust~\cite{scrucca2016mclust} in our paper.

\textbf{Spot visualization.} An essential component of the embedding method, visualisation~\cite{zang2022evnet} allows researchers to intuitively understand and interpret the spatial distribution and relationships between different points. The {\MuST} ensures that the embedding space is accessible to different visualisation techniques, providing clear and meaningful insights. Note that UMAP~\cite{becht2019dimensionality} and t-SNE~\cite{van2008visualizing} may produce false or overly sharp boundaries. We prioritise training an MLP network based on our topology preserving loss to map uniform embeddings onto a two dimensional visualisation space. For details,
\begin{equation}
	\begin{aligned}
		\bm{z}^\text{vi} & = \text{MLP}_{\phi^\text{vi}}(\bm{z}), & \bm{z}^\text{vi} \in \mathbb{R}^2,
	\end{aligned}
\end{equation}
where $\bm{z}^\text{vi}$ is the visualisation embedding of $\bm{z}$ and $\text{MLP}_{\phi^\text{vi}}$ is the multilayer perceptron for the visualisation. The loss function is
\begin{equation}
	\begin{aligned}
		L_\text{vi}(\bm{z}) & = \sum \{
			\mathcal{T}_{ij}  \log(S_{ij}) +(1-\mathcal{T}_{ij})\log(1-S_{ij}) \}, \\
		\mathcal{T}_{ij} & = \kappa(\bm{z}_i, \bm{z}_j), \;\;\; S_{ij} = \kappa \left({\bm{z}}^\text{vi}_i, {\bm{z}}^\text{vi}_j \right)
	\end{aligned}
\end{equation}
where $\bm{z}_i$ and $\bm{z}_j$ are two spot data in the universal latent space $\bm{z}$, $\bm{z}_i^\text{vi}$ and $\bm{z}_j^\text{vi}$ are two spot data in the universal latent space $\bm{z}^\text{vi}$. The $\bm{z}_j$ is the augmented data of $\bm{z}_i$ or sampled from the same modality.

\textbf{Spatial deconvolution.} 
One of the critical challenges in ST is to ensure that the signals detected from a given spot are representative of a single cell type or state, rather than a mixture of multiple sources. The purity of a spot in this context refers to the extent to which the observed transcriptomic profile can be attributed to a single cellular entity. Using the unified embedding space learned by {\MuST}, the profile of each spot is compared to a library of known single cell transcriptomic signatures. The similarity between the spot profile and the single cell signatures provides an estimate of the purity of the spot. A high similarity to a single signature indicates high purity, while similarities to multiple signatures indicate potential mixing. {\MuST} performs this task based on a unified embedding space without the need for reference information. For details, we use the mean vector as the standard vector for each cluster in the universal latent space. The Lasso~\cite{zou2006adaptive} method is then used to determine the sparse weight of the data of each point characterised by the standard vector. The standard deviation of the sparse weight is used to estimate the impurity of the spot.

\textbf{Marker gene analysis with deep learning interpretability.} 
Given the often 'black box' nature of deep learning models, {\MuST} places a strong emphasis on interpretability. This task seeks to understand the model's decision-making process, particularly in the context of identifying marker genes that may be indicative of specific cellular states or activities. In {\MuST}, the amount by which changing a gene changes the position of the embedding in the universal latent space is used to define the importance of that gene. Specifically, changing an important gene will significantly change the position of the embedding of the data in the universal latent space.

\textbf{Spatial trajectories inference.}
Understanding how cells differentiate and develop over time is crucial for many biological applications. {\MuST} facilitates the inference of these trajectories, providing insights into developmental pathways and potential regulatory mechanisms. {\MuST} has integrated the spatial position information into the unified embedding, so we directly use the PAGA~\cite{wolf2019paga} unified embedding method for spatial trajectory inference.

\subsection{Performance Evaluation Protocol}
We evaluate spatial clustering performance using the adjusted rand index (ARI):
\begin{equation}
	\text{ARI} = \frac{\Sigma_{ij}\binom{n_{ij}}{2} - [\Sigma_{i}\binom{a_i}{2}\Sigma_{j}\binom{b_j}{2}] / \binom{n}{2}}{\frac{1}{2}[\Sigma_{i}\binom{a_i}{2}+\Sigma_{j}\binom{b_j}{2}] - [\Sigma_{i}\binom{a_i}{2}\Sigma_{j}\binom{b_j}{2}] / \binom{n}{2}}
\end{equation}
where the ARI near 1 indicates a strong match to ground truth clustering, whereas values near 0 suggest random assignment. In the implementation, we use the adjusted\_rand\_score function from the scikit-learn Python package~\cite{pedregosa2011scikit}.

The mean relative rank error~(MRRE) measures the average of changes in neighbor ranking between the two spaces:
\begin{equation}
	\text{MRRE} = \frac{1} {(M  \frac{|M-2 k|}{k})}\sum_{i=1}^{M} \sum_{j \in
		\mathcal{N}_{i,k}^{(l)}}\frac{|r^{(l)}_{i,j}-r^{(l')}_{i,j}|}{r^{(l)}_{i,j}},
\end{equation}
where $k$ is the sensing range of the metric, $r^{(l)}$ and $r^{(l')}$ indicate the rank of data $j$ to data $i$ in the input space and the embedding space, $M$ is the number of the data.

The value used on gene expression analysis of the new region (Fig.~\ref{fig_DLPFC_151673}G) on DLPFC slide 151673 is calculated as follow:
\begin{equation}
    \bar{x_i} = %
	\left(
		{\text{MEAN}_{i \in \mathcal{N}(i)}(\sum_{g \in \text{Gene}} {x^{tr}_{i, g}})}
	\right)^{0.25}
\end{equation}
where $x^{tr}_{i, g}$ represent the $g$ normalized genes on $i$-th spot of Tra. input $x^{tr}$. The neighbourhood of a spot $\mathcal{N}(i)$ is all spots of its cluster or itself. 

\subsection{Data Description} \label{data description}
To evaluate the performance of {\MuST} and baseline methods in ST, we employ five tissues (eight datasets) corresponding to five subsections in Results section. 
\begin{itemize}
	\item Two mouse embryos (E9.5 and E14.5) datasets are acquired using Stereo-seq data, which we download them from \href{https://db.cngb.org/stomics/mosta/}{https://db.cngb.org/stomics/mosta/}. The E9.5 embryo data consists of $5,913$ bins and $25,568$ genes, while the E14.5 embryo data consists of $92,928$ bins and $18,566$ genes.
	\item The Slide-seqV2 dataset acquired from mouse hippocampus and is downloaded from \href{https://portals.broadinstitute.org/single_cell/study/slide-seq-study}{SCP354}. We use the Puck\_200115\_08 section with $53,172$ spots and $23264$ genes.
	\item The mouse brain tissue is downloaded from the publicly available 10x Genomics Data Repository \href{https://www.10xgenomics.com/resources/datasets}{https://www.10xgenomics.com/resources/datasets}. This dataset has two sections, including a coronal mouse brain section and a mouse sagittal posterior brain section. The former contains $2702$ spots with $32,285$ genes captured and is manually annotated with $52$ regions using the Allen Brain Atlas reference \href{https://mouse.brain-map.org/static/atlas}{https://mouse.brain-map.org/static/atlas}. The latter contains $3355$ spots with $32,285$ genes captured and is manually annotated with $71$ regions using the Allen Brain Atlas reference.
	\item Two mouse olfactory bulb datasets are acquired using Stereo-seq data and Slidev2-seq data, which are further processed and annotated. The former data is downloaded from \href{https://github.com/JinmiaoChenLab/SEDR_analyses}{https://github.com/JinmiaoChenLab/SEDR\_analyses}. This data contains $19,109$ spots and $27,106$ genes. The latter data is downloaded form \href{https://singlecell.broadinstitute.org/single_cell/study/SCP815/highly-sensitive-spatial-transcriptomics-at-near-cellular-resolution-with-slide-seqv2}{SCP815}. This data contains $20,139$ spots and $21,220$ genes. 
	\item The LIBD human dorsolateral prefrontal cortex (DLPFC) with $12$ tissue slices is acquired with 10x Visium. This dataset is available at \href{http://research.libd.org/globus}{http://research.libd.org/globus}. The number of spots in each slice ranges from $3460$ to $4789$, with $33,538$ genes captured. Each slice is manually annotated to contain five to seven regions, namely the DLPFC layers and white matter.
\end{itemize}

\subsection{Comparison with Baseline Methods}
To showcase the effectiveness of {\MuST} in representation learning for ST data, we compare {\MuST} with six state-of-the-art methods, including including GraphST~\cite{long2023spatially}, STAGATE~\cite{dong2022deciphering}, DeepST~\cite{xu2022deepst}, SpaGCN~\cite{hu2021spagcn}, MUSE~\cite{bao2022integrative} and stSME~\cite{pham2020stlearn}.

\textbf{GraphST}. GraphST is a method that integrates graph neural networks with self-supervised contrastive learning specifically for ST data, offering superior performance in spatial clustering, multisample integration, and cell-type deconvolution, outpacing current methods across various tissue types and platforms. For DLPFC dataset, we set the parameter `beta' to 2 and the dimension of the embedding layer to 68. For other datasets, we use the default parameters.

\textbf{STAGATE}. STAGATE is another deep learning model-based method that combines an auto-encoder with a graph attention mechanism to learn latent representation by modeling both gene expression profiles and spatial location information. We ran STAGATE for spatial clustering and vertical and horizontal ST data integration. All experiments are implemented using the recommended parameters in the package vignette. Specifically, with raw gene expressions, the top 3000 highly variable genes are first selected and then log-transformed and normalized according to library size. The parameter `alpha' was set to 0. The learning rate and training epoch are left at the default 0.0001 and 500, respectively.

\textbf{DeepST}. DeepST is an advanced deep learning framework for ST that outperforms existing methods in identifying spatial domains, offering efficient data integration across batches or technologies, and showing adaptability for other spatial omics data. For DLPFC dataset, we set the parameter `nbr\_k' to 5, `tree\_k' to 15, `adj\_w' to 0.1 and `graph\_conv' to ResGatedGraphConv. For other datasets, we use the default parameters.

\textbf{SpaGCN}. SpaGCN is a graph convolutional network approach that integrates gene expression, spatial location information, and histological images for ST data analysis. SpaGCN is one of the only two other methods that can perform horizontal ST data integration. Following the tutorial, we applied SpaGCN to spatial clustering and horizontal ST data integration with the default parameter settings. In particular, the parameter `histology' was set to `False'. The learning rate and max training epoch are set to 0.05 and 200, respectively.

\textbf{MUSE}. MUSE is a novel approach that integrates Mor. and spatially resolved Tra. data to uncover tissue subpopulations, compensates for modality-specific noise, and offers enhanced insights into cellular states and organization in both healthy and diseased tissues. We use the default parameters for all experiments.

\textbf{stSME}. stSME is a comprehensive Python software that employs innovative integrative analysis methods to harness spatial, Mor., and Tra. data from ST, enabling precise cell type identification, modeling of cellular evolution within tissues, and detection of hotspot regions for heightened cell-to-cell interactions in both healthy and diseased contexts. We use the default parameters for all experiments.

\textbf{{\MuST}}. We use dropout to make learning process robust. For all experiments, the input dimension of Tra. modality is 3000 and the input dimension of Mor. modality is 500. For most of experiments, we set the dimension of hidden representation as 72. In information extraction, we employ a graph encoder for each modalities including a structure module (d/input\_dimension, 72, 72, 72). To combine both modalities, we add them by element to get a joint representation.  In information fusion, we employ a MLP for encoding including a structure module (72, 72) for most of datasets. In gene reconstruction, we employ a graph decoder including a structure module(d/72, gene\_dimension).

We run all the experiments using a Ubuntu server and a single A100 GPU with 40GB memory for 10x Visium platform and 80GB memory for other platforms. To ensure that each method achieves its optimal performance, we use the grid search method to find the optimal hyperparameter. For baseline methods, we use the default hyperparameter setting according to the released code.

In the following, we discuss the function of different hyperparameters in {\MuST} and propose typical value ranges.
For Stereo-seq platform, the $K^{tr}$ adjusts the size of Hop-1 neighborhood for Tra. modality and the search space is typically set to $\{7, 9, 10\}$.
The $r_u^{tr}$ adjusts the strength of augmentation which is typically set to $\{0.1, 1\}$. 
The $\nu$ adjusts the degree of freedom of kernel function which is typically set to $\{0.01, 0.05\}$ which is related to dimension.
The $d_{emb}$ denotes the dimension of embedding which is typically set to $\{64, 72\}$.
The $\lambda$ adjusts the effects of the reconstruction loss which is typically set to $\{0.0015, 0.001, 0.015\}$ according to dataset size.
The $\theta$ is set to $\{0.1, 0.2, 1\}$ in accordance with dataset size.
The $\tau$ denotes quantity of the minimum requirement of cells a gene is expressed which is typically set to ${50, 500}$.

For Slide-seq platform, the $K^{tr}$ and $K^{mo}$ adjusts the size of Hop-1 neighborhood for Tra. modality and the search space is typically set to $\{25, 50\}$.
The $\nu$ adjusts the degree of freedom of kernel function which is typically set to $\{0.005, 1\}$ which is related to dimension.
The $d_{emb}$ denotes the dimension of embedding which is typically set to $\{72, 100\}$.
The $\lambda$ adjusts the effects of the reconstruction loss which is typically set to $\{0.00015, 0.005\}$ according to dataset size.

For 10x Visium platform, the $K^{tr}$ and $K^{mo}$ adjusts the size of Hop-1 neighborhood for Tra. modality and Mor. modality and the search space is typically set to $\{3, 5, 7, 9, 11\}$.
The $r_u^{tr}$ and $r_u^{mo}$ adjusts the strength of augmentation for each modality which is typically set to $\{0.1, 0.3, 1\}$. 
The $\nu$ adjusts the degree of freedom of kernel function which is typically set to $\{0.05, 0.13, 0.15\}$ which is related to dimension.
The $\lambda$ adjusts the effects of the reconstruction loss which is typically set to $\{0.001, 0.01, 0.12\}$ according to dataset size.
The $\theta$ adjusts the proportion of Tra. feature which is typically set to $\{0.7, 0.9\}$.
The $N_{MLP}$ adjusts the number of layers in fusion encoder which is typically set to $\{1, 3\}$.
The $lr$ adjusts the learning rate which is typically set to $\{0.001, 0.0014\}$.
We also apply refinement on clustering in specific to DLPFC dataset. We replace the label of each spot with the most occurring label in his 6-nearest neighborhood and itself.

\subsection{Statistics and Reproducibility}
The details about experimental design and statistics used in different data analyses performed in this study are given in the respective sections of results and methods.

\section{Data Availability}
We use publicly available datasets in this study (details in the Sec.~\ref{data description}). To make the results presented in this study reproducible, all processed data are available in Single Cell Portal \href{https://singlecell.broadinstitute.org/single_cell/study/SCP2425/must}{SCP2425} (the access permission will be removed upon acceptance).

\section{Code Availability}
The \MuST\ software package, implemented in Pytorch, is available free from \href{https://github.com/zangzelin/MuST}{https://github.com/zangzelin/MuST} (the access permission will be removed upon acceptance), and as a Supplementary Software 1 accompanying this manuscript.

\section{Acknowledgements}
This work was supported by the National Key R\&D Program of China~(No.2022ZD0115100), the National Natural Science Foundation of China Project~(No. U21A20427), and Project~(No. WU2022A009) from the Center of Synthetic Biology and Integrated Bioengineering of Westlake University. We thank the Westlake University HPC Center for providing computational resources. We thank Yuhao Wang of Westlake University for helping to organize and present the data. We thank Zihan Liu of Westlake University for participating in discussions and providing comments.

\section{Author Contributions}
Stan Z. Li and Zelin Zang proposed this research. Zelin Zang, Liangyu Li, and Yongjie Xu developed the method, Liangyu Li, Yongjie Xu, and Chenrui Duan collected the datasets, and Liangyu Li and Zelin Zang conceived the experiment and wrote the manuscript. Yongjie Xu experimented and analyzed the results. Yi Sun provided biological comments on the manuscript. Kai Wang and Yang You provided valuable suggestions on the manuscript. All authors discussed the results, revised the draft manuscript, and read and approved the final manuscript.

\section{Competing Interests}
The authors declare no competing interests.

\bibliography{mmeg,sample}

\clearpage
\begin{appendices}
\renewcommand{\thefigure}{S\arabic{figure}}
\renewcommand{\thetable}{S\arabic{table}}
\setcounter{figure}{0}
\setcounter{table}{0}

\tableofcontents

\section{Comparing the modality bias phenomenon on input data and {\MuST} representations}
We use the Shaply value~\cite{winter2002shapley} between ground truth and input data/{\MuST} embeddings to evaluate the contributions of transcriptomic (Tra.) modality, morphology (Mor.) modality and location (Loc.) modality. The results show that {\MuST} greatly alleviates the modality bias phenomenon in ST data. In terms of statistical evidence, the contribution Tra. modality, Mor. modality and Loc. modality are more balanced on {\MuST} embeddings (Fig.~\ref{fig_Results_Mouse_Brain_modality_bias}B and Fig.~\ref{fig_Results_Mouse_Embryo_modality_bias}A). In addition, Tra. modality and Mor. modality play different role in terms of tissue structure discovery (Fig.~\ref{fig_Results_Mouse_Brain_modality_bias}C and Fig.~\ref{fig_Results_Mouse_Embryo_modality_bias}B).

\section{Comparing Leiden, Louvain, and mclust clustering on {\MuST} representations}
In {\MuST}, we chose mclust as the default clustering method because our assessment shows that mclust performs better than Leiden and Louvain in most cases. Here, we compare the clustering results on different datasets using Leiden, Louvain, and mclust shown in Fig. \ref{fig_Results_different_clustering_methods}. mclust consistently outperforms Leiden and Louvain on all datastes. Visually, the clusters identified by mclust are more continuous. Nevertheless, we also include Leiden and Louvain in {\MuST} as alternative clustering methods.

\section{Supplementary Tables}
\textbf{Table S1}. Description of all ST datasets used in this study.\\
\textbf{Table S2}. Hyperparameters of {\MuST} in all ST datasets.\\
\textbf{Table S3}. Marker gene selection of {\MuST} on mouse embryo at E14.5 dataset acquired with Stereo-seq.\\
\textbf{Table S4}. Marker gene selection of {\MuST} on mouse embryo at E9.5 dataset acquired with Stereo-seq.\\
\textbf{Table S5}. Marker gene selection of {\MuST} on mouse hippocampus dataset acquired with Slide-seqV2.\\
\textbf{Table S6}. Marker gene selection of {\MuST} on coronal mouse brain section acquired with 10x Visium.\\
\textbf{Table S7}. Marker gene selection of {\MuST} on mouse sagittal posterior brain section acquired with 10x Visium.\\
\textbf{Table S8}. Marker gene selection of {\MuST} on coronal mouse olfactory bulb tissue dataset acquired with Stereo-seq.\\
\textbf{Table S9}. Marker gene selection of {\MuST} on coronal mouse olfactory bulb tissue dataset acquired with Slide-seqV2.\\
\textbf{Table S10}. Marker gene selection of {\MuST} on slice 151673 of DLPFC dataset acquired with 10x Visium.\\

\section{Supplementary Figures}
\textbf{Figure S1}. {\MuST} alleviates the modality bias phenomenon in mouse brain data. \textbf{A} H\&E image of mouse brain coronal section and mouse sagittal posterior brain section. \textbf{B} Contribution of the morphology (Mor.) modality, transcriptome (Tra.) modality and spatial (Spa.) modality input data or MuST embedding to data labeling. \textbf{C} Contribution visualization of the Mor. modality, Tra. modality and Spa. modality input data or MuST embedding to data labeling.\\
\textbf{Figure S2}. {\MuST} alleviates the modality bias phenomenon in mouse embryo data. \textbf{A} Contribution of the morphology (Mor.) modality, transcriptome (Tra.) modality and spatial (Spa.) modality input data or MuST embedding to data labeling on mouse embryo 14.5 and 9.5 datasets. \textbf{C} Contribution visualization of the Mor. modality, Tra. modality and Spa. modality input data or MuST embedding to data labeling.\\
\textbf{Figure S3}. {\MuST} enables accurate identification of different organs in the Stereo-seq mouse embryo. \textbf{A} Tissue domain annotations of the E14.5 mouse embryo data obtained from the original Stereo-seq study. \textbf{B} Clustering results by GraphST and {\MuST} on the E14.5 mouse embryo. \textbf{C} Cluster visualization of selected spatial domains identified by the original Stereo-seq study, {\MuST} and GraphST, respectively. \\
\textbf{Figure S4}. {\MuST} enables accurate identification of different organs in the Stereo-seq mouse embryo. \textbf{A} Tissue domain annotations of the E9.5 mouse embryo data obtained from the original Stereo-seq study. \textbf{B} Clustering results by GraphST and {\MuST} on the E9.5 mouse embryo. \textbf{C} Cluster visualization of selected spatial domains identified by the original Stereo-seq study, {\MuST} and GraphST, respectively. \\
\textbf{Figure S5}. Spatial deconvolution of {\MuST}'s results in adult mouse brain section profiled by 10x Visium. \textbf{A} Allen Brain Institute reference atlas diagram and H\&E image of the mouse cortex. \textbf{B} Allen Brain Institute reference atlas diagram and H\&E image of the mouse sagittal. \textbf{C} Spatial deconvolution of the identified domains by {\MuST} on coronal mouse brain section. \textbf{D} Spatial deconvolution of the identified domains by {\MuST} on mouse sagittal posterior brain section. \textbf{E} Spatial deconvolution of the coronal mouse brain section. \textbf{F} Spatial deconvolution of the mouse sagittal posterior brain section. \\
\textbf{Figure S6}. Embedding analysis of {\MuST} on the mouse olfactory bulb tissue and mouse embryo. \textbf{A} Mean relative rank error (MRRE) scores and UMAP visualizations generated by SpcGCN, MUSE, DeepST, GraphST and {\MuST} representations on the Stereo-seq E14.5 mouse embryo data. Among them, methods MUSE and DeepST have an out of memory (OOM) problem. \textbf{B} MRRE scores and UMAP visualizations generated by the representations of five methods on the Stereo-seq E9.5 mouse embryo data. \textbf{C} MRRE scores, UMAP visualizations and PAGA graphs generated by the representations of five methods on the Stereo-seq mouse olfactory bulb tissue section. \textbf{D} MRRE scores, spot visualizations and PAGA graphs generated by the representations of five methods on the Slide-seqv2 mouse olfactory bulb tissue section. \\
\textbf{Figure S7}. Manual annotations and comparison of spatial domains identified by stSME, SpaGCN, MUSE, DeepST, GraphST and {\MuST} on the 12 slices of DLPFC dataset. \\
\textbf{Figure S8}. Marker genes selection for each anatomical region of mouse hippocampus data acquired with SlideseqV2. \\
\textbf{Figure S9}. Marker genes selection for each region of coronal mouse brain section acquired with 10x Visium. \\
\textbf{Figure S10}. Marker genes selection for each region of mouse sagittal posterior brain section acquired with 10x Visium. \\
\textbf{Figure S11}. Marker genes selection for each laminar organization of coronal mouse olfactory bulb tissue datasets acquired with Slide-seqV2. \\
\textbf{Figure S12}. Marker genes selection for each laminar organization of coronal mouse olfactory bulb tissue datasets acquired with Stereo-seq. \\
\textbf{Figure S13}. Comparison analysis between Leiden, Louvain, and mclust with the output of {\MuST} as input. Visualization of clustering results from Louvain, Leiden, and mclust on Stereo-seq E14.5 mouse embryo data (A), Stereo-seq 9.5 mouse embryo data (B), Stereo-seq mouse olfactory bulb tissue sections (C), Slide-seqV2 mouse olfactory bulb tissue sections (D), 10x Visium coronal mouse brain section (E), 10x Visium mouse sagittal posterior brain section (F), and SlideseqV2 mouse hippocampus data (G). \\

\begin{landscape}
	\begin{table}[ht]
		\renewcommand{\arraystretch}{1.5}
		\caption{Description of all ST datasets used in this study.}\label{tab:dataset_detail}\centering
		\begin{tabular}{l|p{3cm}|p{3cm}llll}
			\toprule
			\multicolumn{1}{l}{Platform} & \multicolumn{1}{p{3cm}}{Tissue}                                              & Section                              & \#Spots/bins & \#Genes & Related figures & Reference                       \\
			\hline
			\multirow{3}{*}{Stereo-Seq}   & \multirow{2}{*}{Mouse embyro}                                                & E14.5                                & 92928      & 18582       & Fig. 2, Fig. C1, Fig. C4, Fig. C11, Tab. C3 & \cite{chen2022spatiotemporal}   \\
			                             &                                                                              & E9.5                                 & 5913       & 25568       & Fig. 2, Fig. C2, Fig. C4, Fig. C11, Tab. C4 & \cite{chen2022spatiotemporal}   \\
			                             & Mouse olfactory bulb                                                         & MOB                                  & 19109      & 27106       &  Fig. 5, Fig. C4, Fig. C9, Fig. C11, Tab. C8             & \cite{chen2022spatiotemporal}                                \\
			\hline
			\multirow{2}{*}{Slide-SeqV2}  & Mouse hippocampus                                                            & Puck\_200115\_08                     & 53172      & 23264       &      Fig. 3, Fig. C6, Fig. C11, Tab. C5           & \cite{rodriques2019slide}       \\
			                             & Mouse olfactory bulb                                                         & Puck\_200127\_15                     & 20139      & 21220       &        Fig. 5, Fig. C4, Fig. C10, Fig. C11, Tab. C9         & \cite{rodriques2019slide}       \\
			\hline
			\multirow{14}{*}{10x Visium} & Mouse brain                                                        & Mouse Brain Section (Coronal)               & 2702       & 32285       &      Fig. 4, Fig. C3, Fig. C7, Fig. C11, Tab. C6           & \cite{10xgeomics}                 \\
			                             & Mouse brain                                               & Mouse Brain Section (Sagittal-Posterior) & 3355       & 32285       &        Fig. 4, Fig. C3, Fig. C8, Fig. C11, Tab. C7         &                                 \\
			                             & \multirow{12}{*}{\parbox{3cm}{Human dorsolateral prefrontal cortex (DLPFC)}} & 151507                               & 4220       & 33538       &    Fig. 6, Fig. C5             & \cite{maynard2021transcriptome} \\
			                             &                                                                              & 151508                               & 4381       & 33538       &    Fig. 6, Fig. C5             & \cite{maynard2021transcriptome} \\
			                             &                                                                              & 151509                               & 4788       & 33538       &    Fig. 6, Fig. C5             & \cite{maynard2021transcriptome} \\
			                             &                                                                              & 151510                               & 4595       & 33538       &    Fig. 6, Fig. C5             & \cite{maynard2021transcriptome} \\
			                             &                                                                              & 151669                               & 3636       & 33538       &     Fig. 6, Fig. C5            & \cite{maynard2021transcriptome} \\
			                             &                                                                              & 151670                               & 3484       & 33538       &     Fig. 6, Fig. C5            & \cite{maynard2021transcriptome} \\
			                             &                                                                              & 151671                               & 4093       & 33538       &     Fig. 6, Fig. C5            & \cite{maynard2021transcriptome} \\
			                             &                                                                              & 151672                               & 3888       & 33538       &     Fig. 6, Fig. C5            & \cite{maynard2021transcriptome} \\
			                             &                                                                              & 151673                               & 3611       & 33538       &    Fig. 6, Fig. C5, Tab. C10             & \cite{maynard2021transcriptome} \\
			                             &                                                                              & 151674                               & 3635       & 33538       &      Fig. 6, Fig. C5           & \cite{maynard2021transcriptome} \\
			                             &                                                                              & 151675                               & 3566       & 33538       &      Fig. 6, Fig. C5           & \cite{maynard2021transcriptome} \\
			                             &                                                                              & 151676                               & 3431       & 33538       &     Fig. 6, Fig. C5            & \cite{maynard2021transcriptome} \\
			\bottomrule
		\end{tabular}
	\end{table}
\end{landscape}

\begin{table}[ht]
	\caption{Hyperparameters of {\MuST} in all ST datasets.}
	\label{tab:hyperparameters}
	\centering
	\begin{tabular}{c|c|c|c|c|c|c|c|c|}
		\toprule
		Section & $K^{mo}$ & $K^{tr}$ & $r^{mo}_{u}$  & $r^{tr}_{u}$ & $\nu$ & $d_{emb}$ & $\theta$ & $\lambda$  \\ \hline

ME95 & 5 & 9 & 0.1 & 1 & 0.05 & 72 & 1 & 0.015 \\
ME145 & 5 & 7 & 0.1 & 1 & 0.05 & 72 & 1 & 0.0015 \\
MOB & 10 & 10 & 0.1 & 0.1 & 0.01 & 64 & 1 & 0.001 \\
DLPFC & 3 & 11 & 1 & 0.3 & 0.13 & 72 & 0.9 & 0.12 \\
Mouse Brain Section (Coronal) & 7 & 7 & 0.1 & 0.1 & 0.05 & 72 & 0.9 & 0.01 \\
Mouse Brain Section (Sagittal-Posterior) & 5 & 9 & 1 & 0.1 & 0.15 & 72 & 0.7 & 0.001 \\
Puck\_200115\_08 & 5 & 25 & 0.1 & 0.5 & 1 & 72 & 1 & 0.005 \\
Puck\_200127\_15 & 10 & 50 & 0.1 & 0.5 & 0.005 & 100 & 1 & 0.00015 \\
		\bottomrule
	\end{tabular}
\end{table}

\begin{table}[ht]
	\caption{Marker gene selection of {\MuST} on mouse embryo at E14.5 dataset acquired with Stereo-seq.}
	\label{tab:gene_selection_ME145}
	\centering
	\begin{tabular}{c|p{4in}}
		\toprule
		Clusters & Marker genes                                                                           \\
		\hline
		C1       & Afp, Hbb-bs, Tuba1a, Ttn, Hbb-bt, Apoa2, Mt1, H19, Col3a1, Col1a2, Lgals1, Fabp7       \\
		C2       & Tuba1a, Fabp7, H19, Col3a1, Hbb-bs, Lgals1, Col1a2, Afp, Stmn2, Hbb-bt, Rtn1, Dbi      \\
		C3       & Col3a1, Tuba1a, Col1a2, Ttn, Hbb-bs, Afp, Myh3, Gphn, Mt1, Hbb-bt, Krt5, Col1a1        \\
		C4       & Afp, Apoa2, Slc25a37, Hbb-bt, Gpx1, Alb, Mt1, Apoa1, Hba-a2, Mt2, Col3a1, Hba-a1       \\
		C5       & Krt5, Tuba1a, Krt15, Mt1, Hbb-bs, Afp, Col3a1, H19, Hbb-bt, Lgals1, Col1a2, Fabp7      \\
		C6       & Ttn, Myh3, Neb, Hbb-bs, Tceal7, Actc1, Myl1, Nrk, Tnni2, Hba-a1, Col3a1, Acta1         \\
		C7       & Ugt2a2, H19, Tuba1a, Col3a1, Sult1e1, Lgals1, Hbb-bs, Afp, Hbb-bt, Col1a2, Fabp7, Igf2 \\
		C8       & Krt5, Cxcl14, Tuba1a, Afp, Perp, Col3a1, Anxa2, Hbb-bs, H19, S100a6, Krt15, Col1a2     \\
		C9       & Col11a1, Col2a1, Tuba1a, Gphn, Hbb-bs, Col3a1, Afp, Malat1, H19, Hbb-bt, Mt1, Col1a2   \\
		C10      & Fabp7, Tuba1a, Slc1a3, Dbi, Hbb-bs, Mylpf, Tubb2b, Spon1, H19, Fez1, Camk1d, Afp       \\
		C11      & Trps1, Alx1, Mt1, Hbb-bs, Hpse2, Tuba1a, Col2a1, Camk1d, Malat1, Msx1, Col3a1, Afp     \\
		C12      & Myh11, Acta2, Mylk, Tuba1a, Actg2, Hbb-bs, Hba-a1, Afp, Prkg1, H19, Col3a1, Fabp7      \\
		C13      & Tuba1a, Map1b, Tmsb10, Col1a2, Col3a1, Hbb-bs, Nnat, Rtn1, H19, Camk1d, Afp, Lgals1    \\
		C14      & Neurod6, Tuba1a, H19, Hbb-bs, Lgals1, Fabp7, Pantr1, Hbb-bt, Actc1, Nfib, Col3a1, Igf2 \\
		C15      & Tuba1a, Rtn1, Hbb-bs, Stmn2, Map1b, H19, Gphn, Gap43, Tubb2a, Tmsb10, Lgals1, Afp      \\
		C16      & Nppa, Myh6, Tuba1a, Tnnt2, Tnni3, H19, Hbb-bs, Afp, Hbb-bt, Actc1, Col3a1, Myl7        \\
		\bottomrule
	\end{tabular}
\end{table}

\begin{table}[ht]
	\caption{Marker gene selection of {\MuST} on mouse embryo at E9.5 dataset acquired with Stereo-seq.}
	\label{tab:gene_selection_ME95}
	\centering
	\begin{tabular}{c|p{4in}}
		\toprule
		Clusters & Marker genes                                                                          \\
		\hline
		C1       & Postn, Fabp7, Igf2, Hbb-bs, H19, Pantr1, Msx1, Gata6, Sox2, Ckb, Hbb-y, Igfbp2        \\
		C2       & Fabp7, Igf2, Hbb-y, Hbb-bh1, Hoxb3os, Ckb, Pantr1, Col3a1, H19, Tmsb4x, Otx2, Cnn2    \\
		C3       & Myl7, Actc1, Acta1, Tnni1, Myl4, Igf2, Fabp7, Tnnt2, Myh7, Hbb-y, Marcks, Lars2       \\
		C4       & Fabp7, Igf2, Hbb-bh1, Ckb, H19, Hbb-y, Dlk1, Col3a1, Tmsb4x, Crabp1, Rfx4, Tuba1a     \\
		C5       & Igf2, Fabp7, Myl7, Ckb, Zic1, Mest, H19, Crabp1, Foxf1, Peg3, Hbb-y, Actc1            \\
		C6       & Fabp7, Igf2, Ckb, Hbb-bh1, H19, Hbb-y, Cnn2, Pantr1, Col3a1, Rfx4, Sox2, Vim          \\
		C7       & Lars2, Fabp7, Marcks, Igf2, Cnn2, Flrt2, Zc3h7a, Ckb, Ebf1, Gm26561, Sulf1, Acta1     \\
		C8       & Meox1, Myl7, Fabp7, Igf2, Msx1, Dmrt2, Mylpf, Acta1, Lars2, Actc1, Zc3h7a, Pantr1     \\
		C9       & Fabp7, Igf2, Pantr1, Sox2, Id4, Hbb-y, Ckb, Rfx4, H19, Dlk1, Col3a1, Cnn2             \\
		C10      & Fabp7, Igf2, Vim, Ckb, Hbb-y, Pantr1, Rfx4, Sox2, H19, Msi1, Lars2, Marcks            \\
		C11      & Cpox, Hbb-bt, Fabp7, Mt2, Igf2, H19, Hba-a2, Mt1, Hbb-y, Sox2, Abcb10, Pantr1         \\
		C12      & Fabp7, Igf2, Hbb-y, Ckb, H19, Crabp1, Pantr1, Zic1, Hbb-bh1, Sox2, Myl7, Vim          \\
		C13      & Mylpf, Fabp7, Igf2, Pantr1, Hba-x, Hbb-bh1, Vim, Myl7, Meox2, Ckb, H19, Nnat          \\
		C14      & Igf2, Fabp7, Ckb, Hbb-bh1, Hbb-y, H19, Ntn1, Msx1, Tmsb4x, Mest, Nkx6-1, Col3a1       \\
		C15      & Fabp7, Igf2, Msx1, H19, Pantr1, Acta1, Hand2, Rfx4, Actc1, Tnni1, Dlk1, Myl7          \\
		C16      & Hbb-y, Fabp7, Igf2, Sox2, H19, Gm42418, Ckb, Pantr1, Bpgm, Mt2, Cenpa, Ldha           \\
		C17      & Fabp7, Igf2, H19, Myl7, Hbb-bs, Ckb, Pantr1, Dlk1, Sox2, Postn, Mest, Hoxc9           \\
		C18      & Pax1, Fabp7, Arg1, Meox2, Vim, Igf2, Ckb, Sox2, Myl7, Pantr1, H19, Vcan               \\
		C19      & Fabp7, Igf2, Ebf1, Pantr1, Ckb, Stmn2, Hbb-y, H19, Hbb-bh1, Lars2, Nefm, Vim          \\
		C20      & Afp, Fabp7, Igf2, Pantr1, Ckb, H19, Myl7, Hbb-y, Sox2, Lars2, Actc1, Vim              \\
		C21       & Fabp7, Hbb-bh1, Igf2, Hbb-y, Hoxb3os, Col3a1, Pantr1, Ckb, Tmsb4x, H19, Crabp1, Lars2 \\
		C22       & Igf2, Fabp7, Ckb, Hbb-y, H19, Ntn1, Hbb-bh1, Col3a1, Rfx4, Cnn2, Pantr1, Tmsb4x       \\
		\bottomrule
	\end{tabular}
\end{table}

\begin{table}[ht]
	\caption{Marker gene selection of {\MuST} on mouse hippocampus dataset acquired with Slide-seqV2.}
	\label{tab:gene_selection_hip}
	\centering
	\begin{tabular}{c|p{4in}}
		\toprule
		Clusters           & Marker genes                                                                          \\
		\hline
		C1                 & Plp1, Pcp4, Snap25, Mbp, Meg3, Nrgn, Hpca, Ncald, Mef2c, Mobp, Ttr, Ppp3ca            \\
		C2                 & Pcp4, Hpca, Nrgn, Plp1, Rora, Pcp4l1, Mbp, Ccdc136, Cit, Ttr, Tcf7l2, Tnnt1           \\
		C3                 & Nrgn, Snap25, Mef2c, Pcp4, Ncald, Hpca, Plp1, Meg3, Lamp5, Pde1a, 3110035E14Rik, Lmo4 \\
		C4                 & Plp1, Mbp, Pcp4, Mobp, Hpca, Qdpr, Mal, Cnp, Cldn11, Nrgn, Sept4, Ptgds               \\
		C5 (CA1)           & Hpca, Ppp3ca, Nrgn, Itpka, Wipf3, Prkcb, Plp1, Tmsb4x, Pcp4, Cck, Meg3, Mbp           \\
		C6 (CA3)           & Hpca, Cpne6, Ppp3ca, Cplx2, Snap25, Plp1, Chgb, Hs3st4, Nrgn, Pcp4, Calm2, Tmsb4x     \\
		C7 (Dentate Gyrus) & Hpca, Ppp3ca, Ncdn, Nrgn, Plp1, Cplx2, Snap25, Pcp4, Olfm1, Mbp, Itpka, Meg3          \\
		C8 (Interneuron)   & Gad1, Sst, Pcp4, Plp1, Hpca, Mbp, Gad2, Nrgn, Cnr1, Snap25, Meg3, Ncald               \\
		C9                 & Meg3, Nrgn, Snhg11, Pcp4, Snap25, Hpca, Plp1, Malat1, Calb2, Ptk2b, Mef2c, Tmsb4x     \\
		C10 (LH)           & Pcp4, Nrgn, Hpca, Plp1, Ptgds, Nwd2, Snap25, Nnat, Nefm, Ttr, Gap43, Meg3             \\
		C11 (V3)           & Ttr, Nrgn, Pcp4, Plp1, 1500015O10Rik, Hpca, Zic1, Pcp4l1, Malat1, Mbp, Meg3, Enpp2    \\
		C12 (MH)           & Nwd2, Zic1, Ttr, Hpca, Calb2, Pcp4, Tac2, Cd63, Plp1, Tcf7l2, Meg3, Gng8              \\
		C13                & Apod, Ptgds, Aqp4, Agt, Pcp4, Vim, Hpca, Plp1, Nrgn, Zic1, Nwd2, Ttr                  \\
		\bottomrule
	\end{tabular}
\end{table}

\begin{table}[ht]
	\caption{Marker gene selection of {\MuST} on coronal mouse brain section acquired with 10x Visium.}
	\label{tab:gene_selection_cmbs}
	\centering
	\begin{tabular}{c|p{4in}}
		\toprule
		Clusters                & Marker genes                                                                                 \\
		\hline
		C1                      & Tcf7l2, Nrgn, Sparc, Ddn, Ctxn3, Camk2n1, Slc17a6, Slc17a7, Hlf, Ctxn1, Rims3, Adarb1        \\
		C2                      & Nrgn, Mobp, Tcf7l2, Slc17a7, Mbp, Pvalb, Sparc, Prkcd, Vxn, Camk2n1, Fxyd7, Ddn              \\
		C3                      & Hap1, Baiap3, Pmch, 6330403K07Rik, Slc17a7, Nrgn, Camk2n1, Resp18, Nsmf, Tcf7l2, Ddn, Cartpt \\
		C4 (V3)                 & Slc17a7, Mia, Ncdn, Thy1, Camk2n1, Hap1, Cck, Mbp, Nrgn, Baiap3, Tcf7l2, Nsmf                \\
		C5                      & Ddn, Ttr, Camk2n1, Fth1, Hpca, Nrgn, Slc17a7, Mbp, Spink8, Pantr1, Cabp7, Ctxn1              \\
		C6 (Dentate Gyrus)      & Camk2n1, Slc17a7, Nrgn, Vxn, Tcf7l2, Eef1a1, Pcp4, Ddn, Hpca, Nsmf, Mbp, Adcy1               \\
		C7                      & Syn2, Camk2n1, Nptxr, C1ql3, Nr2f2, Vxn, Ddn, Nnat, Uchl1, Nrgn, Lypd1, Pcp4                 \\
		C8                      & Prkcd, Adarb1, Tcf7l2, Nrgn, Ctxn1, Rora, Tnnt1, Pcp4, Atp2b1, Slc17a7, Ptpn3, Mobp          \\
		C9 (Cerebral cortex 2)  & Camk2n1, Nrgn, Mef2c, Snap25, Dkkl1, Mbp, Plcxd2, Slc17a7, Atp1a1, Pcp4, Ddn, Vxn            \\
		C10                     & Camk2n1, Ddn, Slc17a7, Mbp, Penk, Hpcal4, Hpcal1, Hpca, Fth1, Syn2, Ctxn1, Vxn               \\
		C11                     & Slc17a7, Nrgn, Tcf7l2, Camk2n1, Sparc, Ddn, Baiap3, Nnat, Rpl17, Hap1, Tmsb4x, Thy1          \\
		C12                     & Camk2n1, Ddn, Mbp, Olfm1, Hpca, Nrgn, Slc17a7, Nptxr, Penk, Ctxn1, Hpcal4, Vxn               \\
		C13                     & Penk, Slc17a7, Camk2n1, Ctxn1, Arpp21, Hpcal4, Nrgn, Ddn, Tcf7l2, Mbp, Wfs1, Eef1a1          \\
		C14 (Cerebral cortex 1) & Camk2n1, Mbp, Pcp4, Nrgn, Lamp5, Tubb2a, Cplx2, Tmsb10, Ddn, Atp1a1, Slc17a7, Vxn            \\
		C15 (Cerebral cortex 3) & Camk2n1, Slc17a7, Vxn, Eef1a1, Ddn, Tcf7l2, Mbp, Nrgn, 1110008P14Rik, Fth1, Nptxr, Ttr       \\
		C16 (Fiber tracts)      & Mbp, Fth1, Mobp, Thy1, Ncdn, Camk2n1, Tcf7l2, Nrgn, Tmsb4x, Slc17a7, Snap25, Cnp             \\
		C17                     & Prkcd, Adarb1, Tcf7l2, Rora, Atp2b1, Pcp4, Nrgn, Tnnt1, Camk2n1, Ctxn1, Slc17a7, Ptpn3       \\
		C18 (CA1)               & Hpca, Camk2n1, Nrgn, Tmsb4x, Wipf3, Slc17a7, Vxn, Spink8, Tcf7l2, Ddn, Mbp, Ppp3ca           \\
		C19 (Cerebral cortex 4) & Vxn, Tmsb4x, Slc17a7, Camk2n1, Mbp, Diras2, Ttc9b, Mobp, Nr4a2, Nptxr, Ncald, Nrgn           \\
		C20 (CA3)               & Camk2n1, Slc17a7, Tcf7l2, Nrgn, Vxn, Hpca, Ctxn1, Mbp, Ddn, Fth1, Hpcal4, Cnih2              \\
		\bottomrule
	\end{tabular}
\end{table}

\begin{table}[ht]
	\caption{Marker gene selection of {\MuST} on the mouse sagittal posterior brain section acquired with 10x Visium.}
	\label{tab:gene_selection_mpbs}
	\centering
	\begin{tabular}{c|p{4in}}
		\toprule
		Clusters                & Marker genes                                                                           \\
		\hline
		C1 (CA1)                & Mbp, Pcp2, Camk2a, Nrgn, Hpca, Ddn, Rasgrp1, Cck, Nnat, Stmn1, Pvalb, Ctxn1            \\
		C2                      & Nrgn, Mbp, Ddn, Camk2a, Pcp4, Slc1a2, Pvalb, Ctxn1, Pcp2, Plp1, Pantr1, Snap25         \\
		C3                      & Mbp, Camk2a, Cpe, Rasgrp1, Plp1, Atp1a1, Nnat, Ctxn1, Slc17a7, Snap25, Nrgn, Cck       \\
		C4                      & Nnat, Ccn3, Ly6h, Mbp, Dcn, Crym, Nptxr, Ctxn1, Ttr, Ddn, Trh, Pcp2                    \\
		C5 (Coronal structure)  & Atp1a1, Camk2a, Grin2c, Nrep, Adcy1, Mbp, Pcp2, Cbln3, Pvalb, Cbln1, Gabra6, Nrgn      \\
		C6 (Fiber tracts)       & Mbp, Plp1, Nrgn, Cpe, Cck, Atp1a1, Camk2a, Trf, Mobp, Pcp2, Camkv, Ctxn1               \\
		C7 (Cerebral cortex 2)  & Pcp2, Mbp, Nrgn, Atp1a1, Camk2n1, Pcp4, Camk2a, Slc1a2, 1110008P14Rik, Nnat, Dkk3, Ddn \\
		C8                      & Mbp, Cpe, Pvalb, Nefl, Mobp, Snap25, Pcp2, Vamp1, Camk2a, Nrgn, Cck, Hoxb5             \\
		C9 (Dentate Gyrus)      & Mbp, Pcp2, Ddn, Camk2a, Olfm1, Pvalb, Nrgn, Synpr, Snhg11, Ctxn1, Ncdn, Pcp4           \\
		C10 (Thin layer 1)      & Pcp2, Mbp, Pvalb, Atp1a1, Itpr1, Gng13, Camk2a, Mobp, Stmn1, Cbln1, Car8, Cbln3        \\
		C11 (Thin layer 2)      & Pcp2, Mbp, Nptx1, Pvalb, Nrgn, Cbln1, Calm2, Ppp1r17, Car8, Gabra6, Camk2a, Pcp4       \\
		C12 (Thin layer 3)      & Pcp2, Mbp, Cbln1, Pvalb, Atp1a1, Mobp, Camk2a, Nptx1, Car8, Gabra6, Cbln3, Calm2       \\
		C13 (Cervical)          & Hoxb5, Mbp, Pcp2, Cck, Pvalb, Camk2a, Slc6a5, Nefm, Vamp1, Nrgn, Zwint, Vsnl1          \\
		C14                     & Mbp, Camk2n1, Camk2a, Mef2c, Nnat, Cck, Ctxn1, Pcp2, Lypd1, Stx1a, Atp1a1, Snhg11      \\
		C15 (Cerebral cortex 1) & Prkcd, Tnnt1, Cck, Pcp2, Rasgrp1, Ntng1, Mbp, Rgs16, Tcf7l2, Pdp1, Nrgn, Ndrg4         \\
		C16                     & Ddn, Pvalb, Camk2a, Mbp, Nrgn, Vsnl1, C1ql2, Uchl1, Atp1a1, Pcp2, Ctxn1, Stmn1         \\
		C17                     & Mbp, Pcp2, Nrgn, Camk2a, Pvalb, Igfbp2, Nnat, Ddn, Cck, Ctxn1, Atp1a1, Ccn3            \\
		C18                     & Mbp, Nptxr, Trh, Pcp2, Nrgn, Ctxn1, Nnat, Camk2a, Ccn3, Lypd1, Hap1, Ddn               \\
		C19                     & Lypd1, Ctxn1, Pgrmc1, Mbp, Nnat, Ccn3, Ttr, Nr2f2, Pcp2, Ptgds, Ly6h, Trh              \\
		C20 (CA3)               & Crym, Hpca, Pvalb, Mbp, Camk2a, Nrgn, Cpne6, Ccn3, Pcp2, Ctxn1, Ak5, Cnih2             \\
		\bottomrule
	\end{tabular}
\end{table}

\begin{table}[ht]
	\caption{Marker gene selection of {\MuST} on coronal mouse olfactory bulb tissue dataset acquired with Stereo-seq.}
	\label{tab:gene_selection_mob_stereo}
	\centering
	\begin{tabular}{c|p{4in}}
		\toprule
		Clusters & Marker genes                                                                    \\
		\hline
		ONL      & Pcp4, Gad1, Gng13, Fabp7, Ptn, S100a5, Calb2, Nrxn3, Apoe, Kcnb2, Ptgds, Ptprd  \\
		EPL      & Calb2, Nrsn1, Pcp4, Nxph1, Nrxn3, Apoe, Gng13, Fabp7, Ptn, S100a5, Cck, Zic1    \\
		GCL      & Gad1, Pcp4, Pcp4l1, Nme7, Ncdn, Ptprd, Stxbp6, Spp1, Gabra1, Atp1b1, Prkca      \\
		IPL      & Spp1, Stmn2, Uchl1, Gad1, Gabra1, Pcp4, Cpe, Lhfpl3, Nmb, Nef, Ptprd, Cplx1     \\
		MCL      & Gad1, Pcp4, Nrxn3, Kcnb2, Ppp3ca, Atp1a2, Plp1, Calb2, Nme7, Pbx3, Tenm3, Cpne4 \\
		GL       & Gng13, Ptn, Fabp7, Ptgds, S100a5, Apoe, Atp1a2, Npy, Plp1, Gad1, Pcp4, Snap25   \\
		RMS      & Plp1, Pcp4, Nrep, Gad1, Nme7, Nrxn3, Pbx3, Mobp, Ptgds, Apoe, Tenm3, Calb2      \\
		\bottomrule
	\end{tabular}
	\begin{tablenotes}
		\item[*] Here we assign clusters to the corresponding tissue structure: olfactory nerve layer (ONL), external plexiform layer (EPL), granule cell layer (GCL), internal plexiform layer (IPL), mitral cell layer (MCL), glomerular layer (GL), and rostral migratory stream (RMS) according to the ground truth.
	\end{tablenotes}
\end{table}

\begin{table}[ht]
	\caption{Marker gene selection of {\MuST} on coronal mouse olfactory bulb tissue dataset acquired with Slide-seqV2.}
	\label{tab:gene_selection_mob_slide}
	\centering
	\begin{tabular}{c|p{4in}}
		\toprule
		Clusters & Marker genes                                                                       \\
		\hline
		GCL      & Pcp4, Nrxn3, Camk2b, Doc2g, Meis2, Ppp3ca, Nrsn1, S100a5, Cck, Fabp7, Atp1a2, Dlg2 \\
		EPL      & Pcp4, S100a5, Nrxn3, Doc2g, Camk2b, Fabp7, Apod, Cck, Atp1a2, Ppp3ca, Meis2, Nrsn1 \\
		MCL      & Doc2g, Cck, Rab3b, Pcp4, Olfm1, S100a5, Nrsn1, Map1b, Cdhr1, Slc17a7, Fabp7, Nrxn3 \\
		ONL      & S100a5, Fabp7, Apod, Pcp4, Ptgds, Ptn, Gng13, Gad1, Nrsn1, Omp, Doc2g, Plp1        \\
		RMS      & Pcp4, Mbp, Nrxn3, Doc2g, Nrgn, S100a5, Tubb2b, Camk2b, Meis2, Nrsn1, Nrep, Mobp    \\
		GL       & S100a5, Nrsn1, Pcp4, Fabp7, Apod, Calb2, Gng13, Ptgds, Cck, Nrxn3, Doc2g, Vsnl1    \\
		IPL      & Pcp4, Atp1b1, Nrxn3, Doc2g, Nrsn1, Camk2b, Ppp3ca, Gad1, Scg2, Cck, Rab3b, Fxyd6   \\
		AOBgr    & Nrxn3, Cpne6, Pcp4, Tac1, Doc2g, Fxyd6, Calb2, Plcb1, Camk2b, Dpp6, Jph4, Atp2b4   \\
		AOB      & Fxyd6, Slc17a7, Rab3b, Pcp4, Stmn2, Uchl1, Doc2g, Cdhr1, Calb2, Cck, Nrxn3, Ptprd  \\
		\bottomrule
	\end{tabular}
	\begin{tablenotes}
		\item[*] Here we assign clusters to the corresponding tissue structure: granule cell layer (GCL), external plexiform layer (EPL), mitral cell layer (MCL), olfactory nerve layer (ONL), rostral migratory stream (RMS), glomerular layer (GL), internal plexiform layer (IPL), granular layer of the accessory olfactory bulb (AOBgr) and accessory olfactory bulb (AOB) according to the ground truth.
	\end{tablenotes}
\end{table}

\begin{table}[ht]
	\caption{Marker gene selection of {\MuST} on slice 151673 of DLPFC dataset acquired with 10x Visium.}
	\label{tab:gene_selection_151673}
	\centering
	\begin{tabular}{c|p{4in}}
		\toprule
		Clusters     & Marker genes                                                           \\
		\hline
		White Matter & MBP, PLP1, GFAP, CNP, PTGDS, TF, SNAP25, CRYAB, TMSB10, MOBP           \\
		Layer 1      & MBP, GFAP, SNAP25, PLP1, MALAT1, TUBA1B, NEFL, CAMK2N1, STMN2, SLC24A2 \\
		Layer 2      & MBP, CAMK2N1, TUBA1B, HPCAL1, NEFL, MALAT1, PLP1, TMSB10, ENC1, SOWAHA \\
		Layer 3      & MBP, PLP1, TMSB10, DIRAS2, GFAP, CAMK2N1, TUBA1B, NEFL, SNAP25, MALAT1 \\
		Layer 4      & MBP, PLP1, TUBA1B, IGKC, IGLC2, GFAP, SNAP25, TMSB10, MALAT1, SCGB1D2  \\
		Layer 5      & MBP, TMSB10, PCP4, PLP1, SNAP25, GFAP, TUBA1B, DIRAS2, HPCAL1, MALAT1  \\
		Layer 6      & MBP, PLP1, DIRAS2, GFAP, PCP4, TMSB10, KRT17, SNCG, B3GALT2, SNAP25    \\
		\bottomrule
	\end{tabular}
\end{table}

\begin{figure}[ht]
    \centering
    \includegraphics[width=\textwidth]{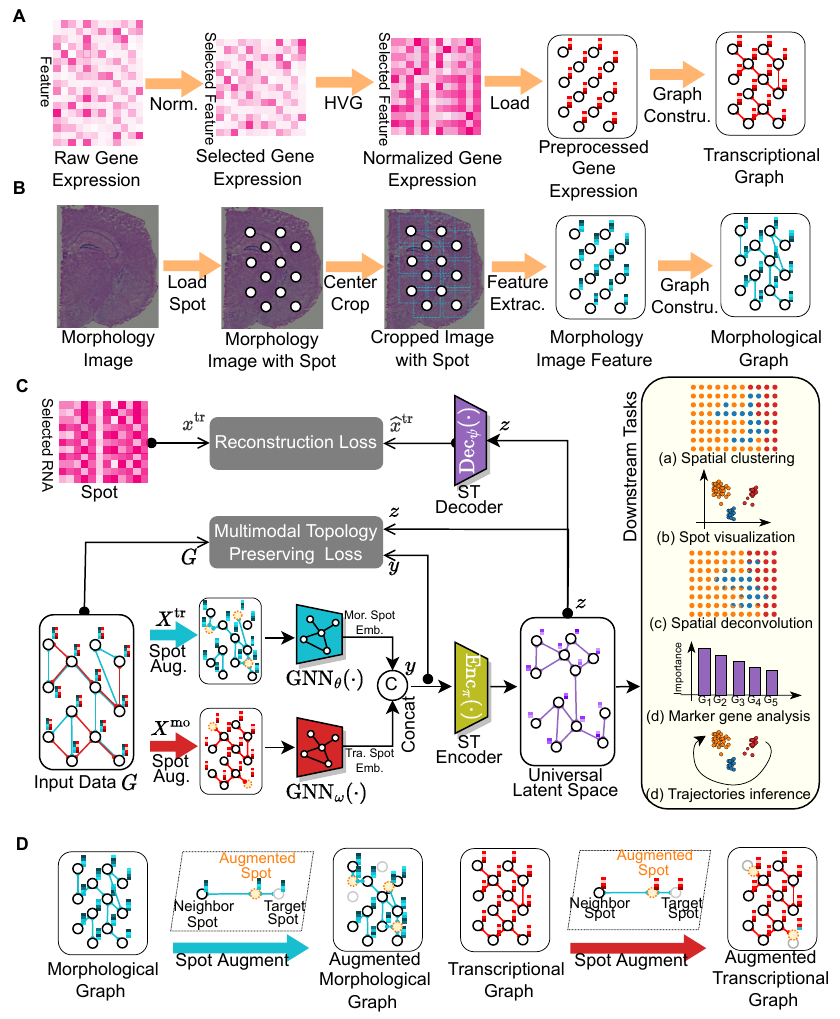}
    \caption{
		Details framework of MuST. 
		\textbf{A.} Preprocessing for transcriptomics modality. The gene expression counts are log transformed and normalised and then we select the 3000 most variable genes.
		\textbf{B.} Preprocessing for morphology modality. We centre and crop a $224 \times 224$ pixel area around the probe coordinates. These images are then processed by the ResNet50 architecture~\cite{kaiming2016resnet}, to distill a 2048-dimensional feature vector and use PCA map the 2048-dimensional to 50-dimensional by PCA. 
		\textbf{C.} Neural network data processing flow.
		\textbf{D.} Data augmentation of morphology and transcriptional.
	}
    \label{fig_method_framework}
\end{figure}

\begin{figure}[ht]
	\centering
	\includegraphics[width=5.5in]{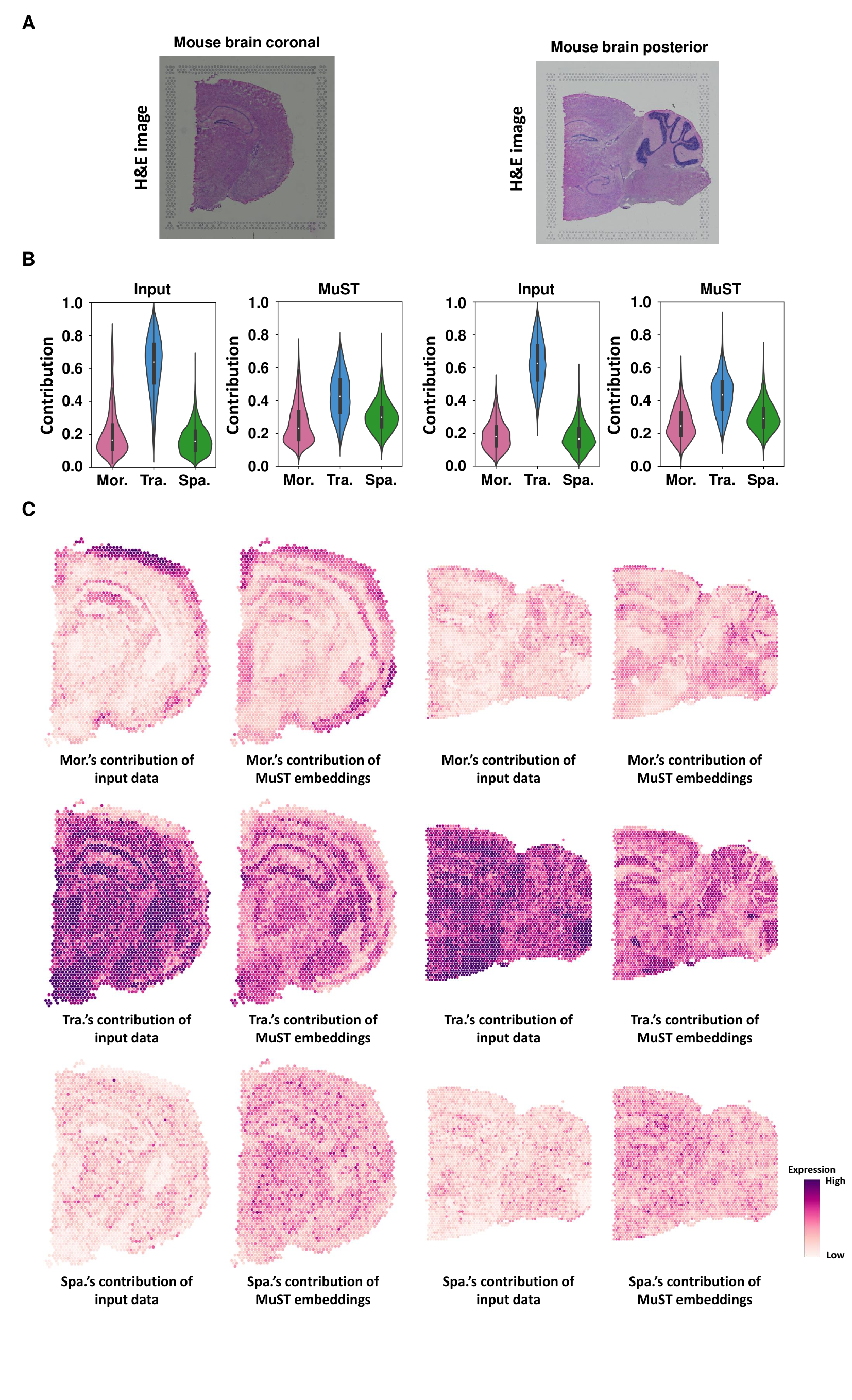}
	\caption{\textbf{{\MuST} alleviates the modality bias phenomenon in mouse brain data. A} H\&E image of mouse brain coronal section and mouse sagittal posterior brain section. \textbf{B} Contribution of the morphology (Mor.) modality, transcriptome (Tra.) modality and spatial (Spa.) modality input data or MuST embedding to data labeling. \textbf{C} Contribution visualization of the Mor. modality, Tra. modality and Spa. modality input data or MuST embedding to data labeling.}
	\label{fig_Results_Mouse_Brain_modality_bias}
\end{figure}

\begin{figure}[ht]
	\centering
	\includegraphics[width=5.5in]{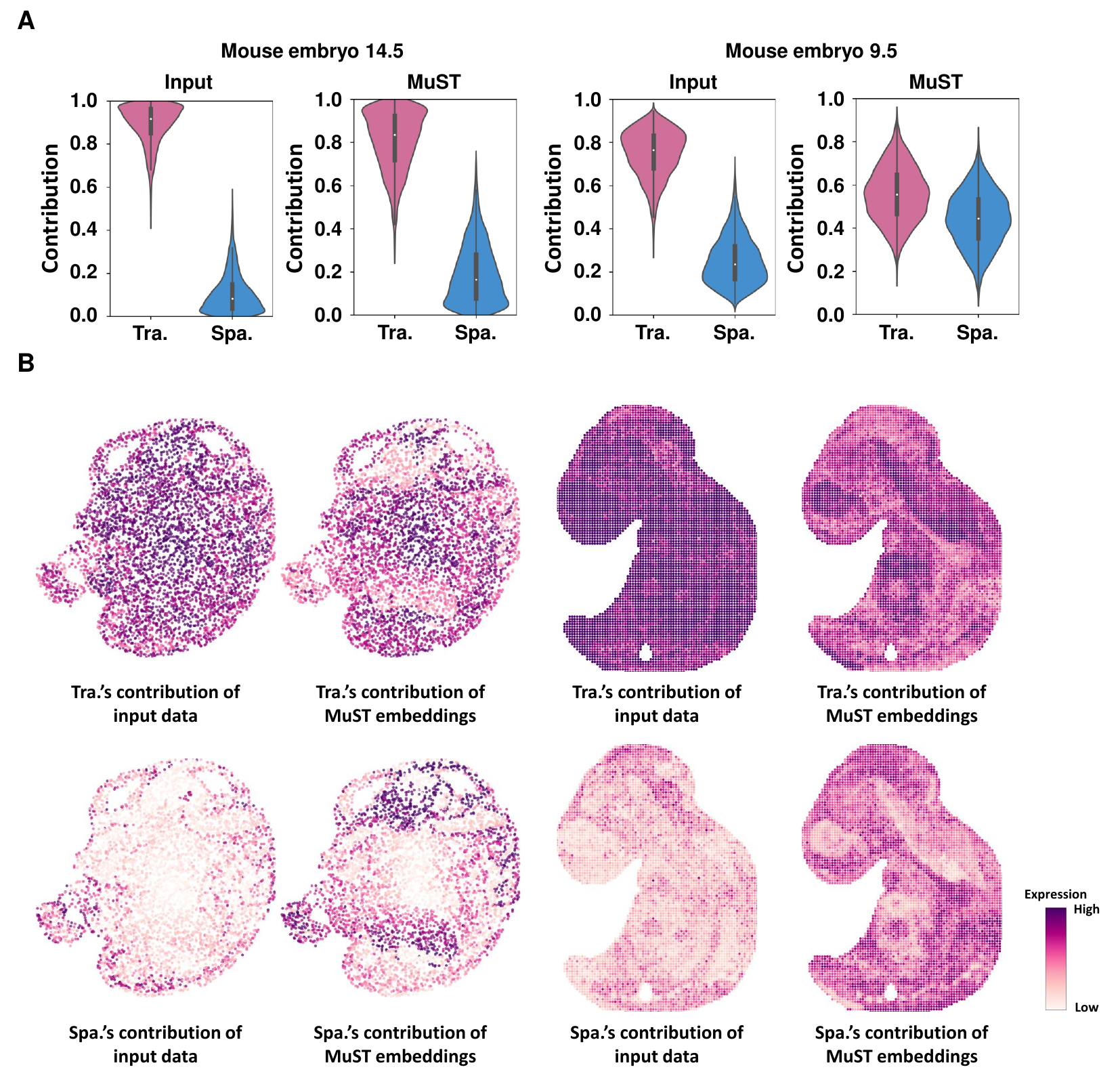}
	\caption{\textbf{{\MuST} alleviates the modality bias phenomenon in mouse embryo data. A} Contribution of the morphology (Mor.) modality, transcriptome (Tra.) modality and spatial (Spa.) modality input data or MuST embedding to data labeling on mouse embryo 14.5 and 9.5 datasets. \textbf{C} Contribution visualization of the Mor. modality, Tra. modality and Spa. modality input data or MuST embedding to data labeling.}
	\label{fig_Results_Mouse_Embryo_modality_bias}
\end{figure}

\begin{figure}[ht]
	\centering
	\includegraphics[width=5.5in]{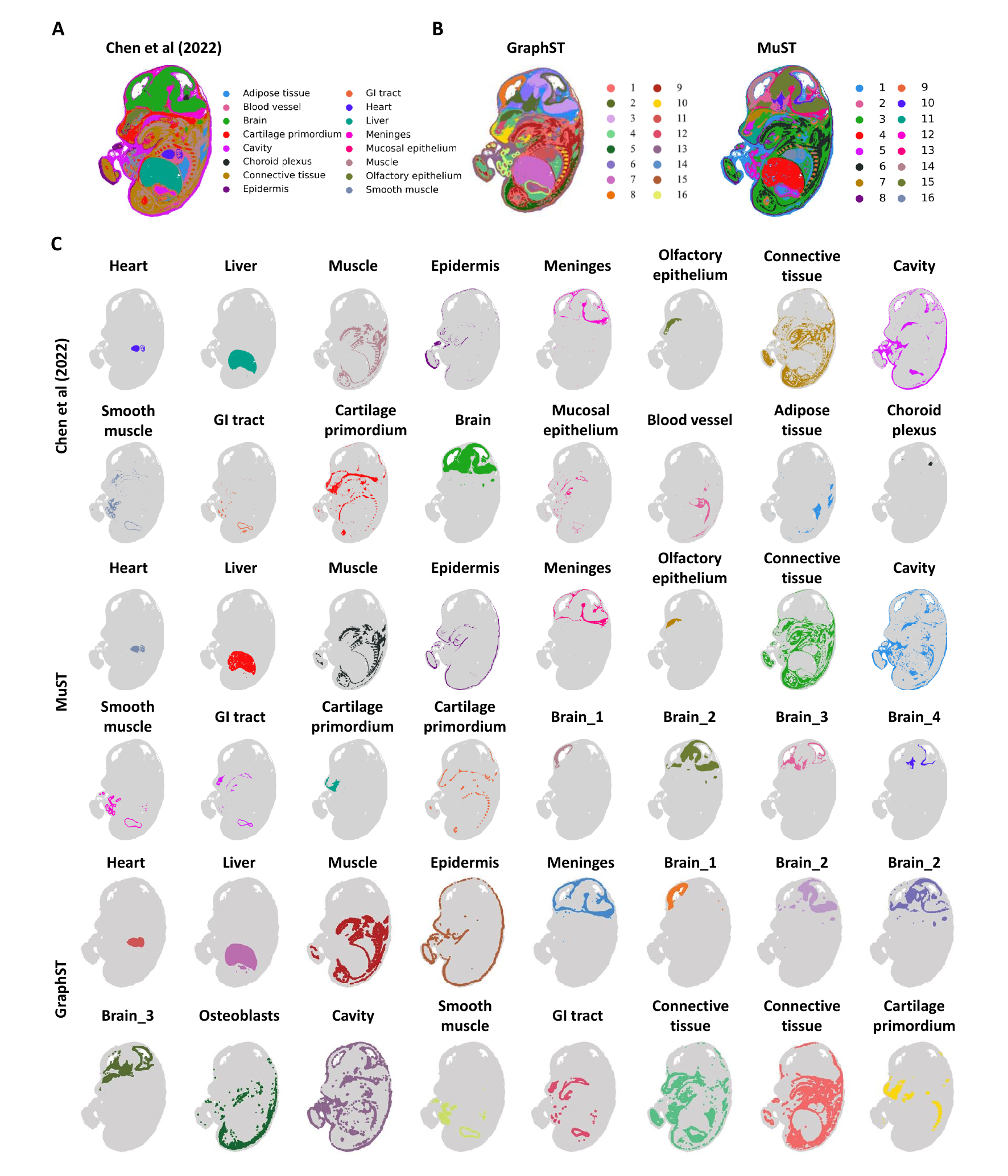}
	\caption{\textbf{{\MuST} enables accurate identification of different organs in the \href{http://research.libd.org/spatialLIBD/}{Stereo-seq} mouse embryo. A} Tissue domain annotations of the E14.5 mouse embryo data obtained from the original Stereo-seq study. \textbf{B} Clustering results by GraphST and {\MuST} on the E14.5 mouse embryo. \textbf{C} Cluster visualization of selected spatial domains identified by the original Stereo-seq study, {\MuST} and GraphST, respectively.}
	\label{fig_Results_Mouse_Embryos_145}
\end{figure}

\begin{figure}[ht]
	\centering
	\includegraphics[width=5.5in]{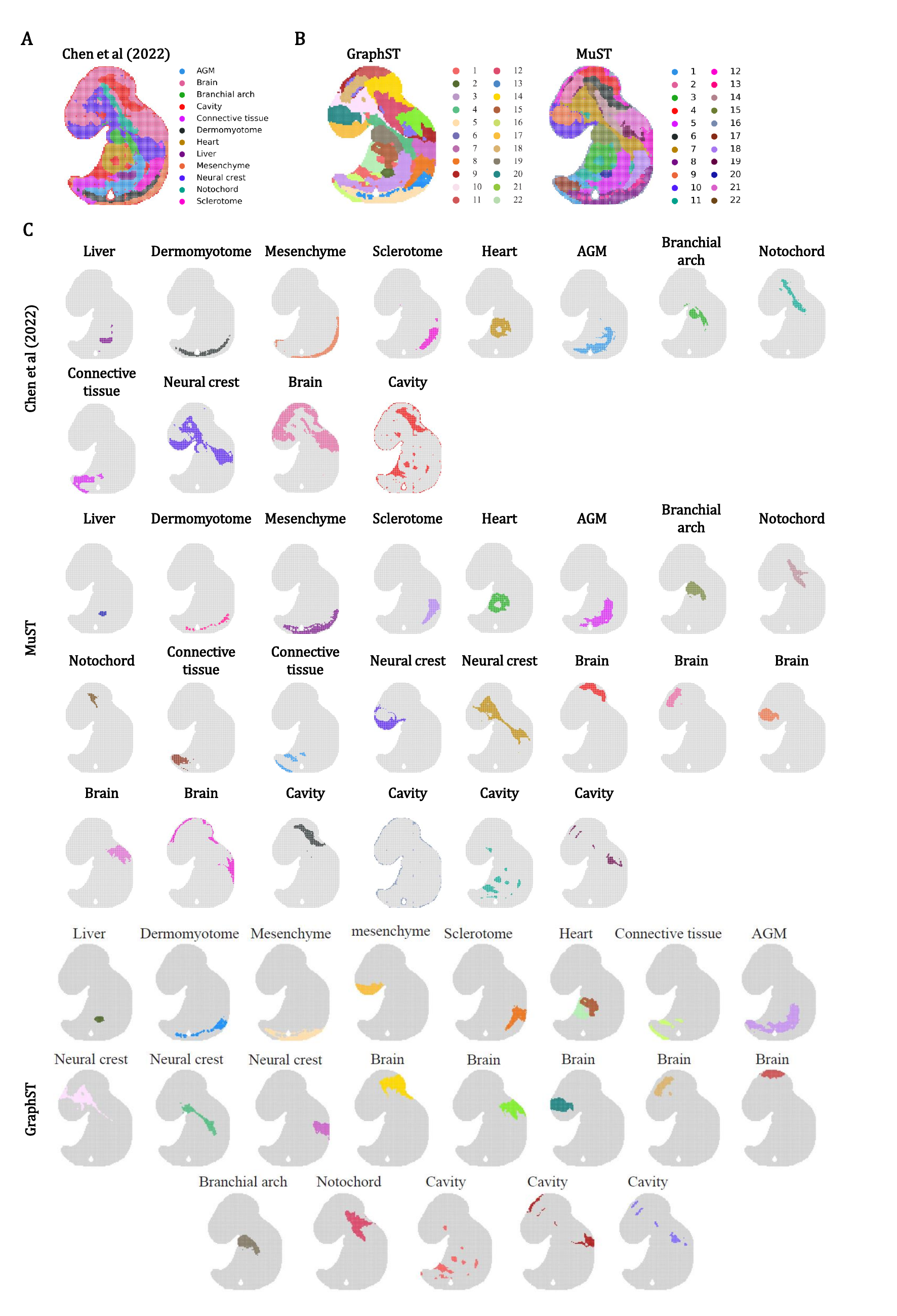}
	\caption{\textbf{{\MuST} enables accurate identification of different organs in the \href{http://research.libd.org/spatialLIBD/}{Stereo-seq} mouse embryo. A} Tissue domain annotations of the E9.5 mouse embryo data obtained from the original Stereo-seq study. \textbf{B} Clustering results by GraphST and {\MuST} on the E9.5 mouse embryo. \textbf{C} Cluster visualization of selected spatial domains identified by the original Stereo-seq study, {\MuST} and GraphST, respectively.}
	\label{fig_Results_Mouse_Embryos_95}
\end{figure}

\begin{figure}[ht]
	\centering
	\includegraphics[width=\textwidth]{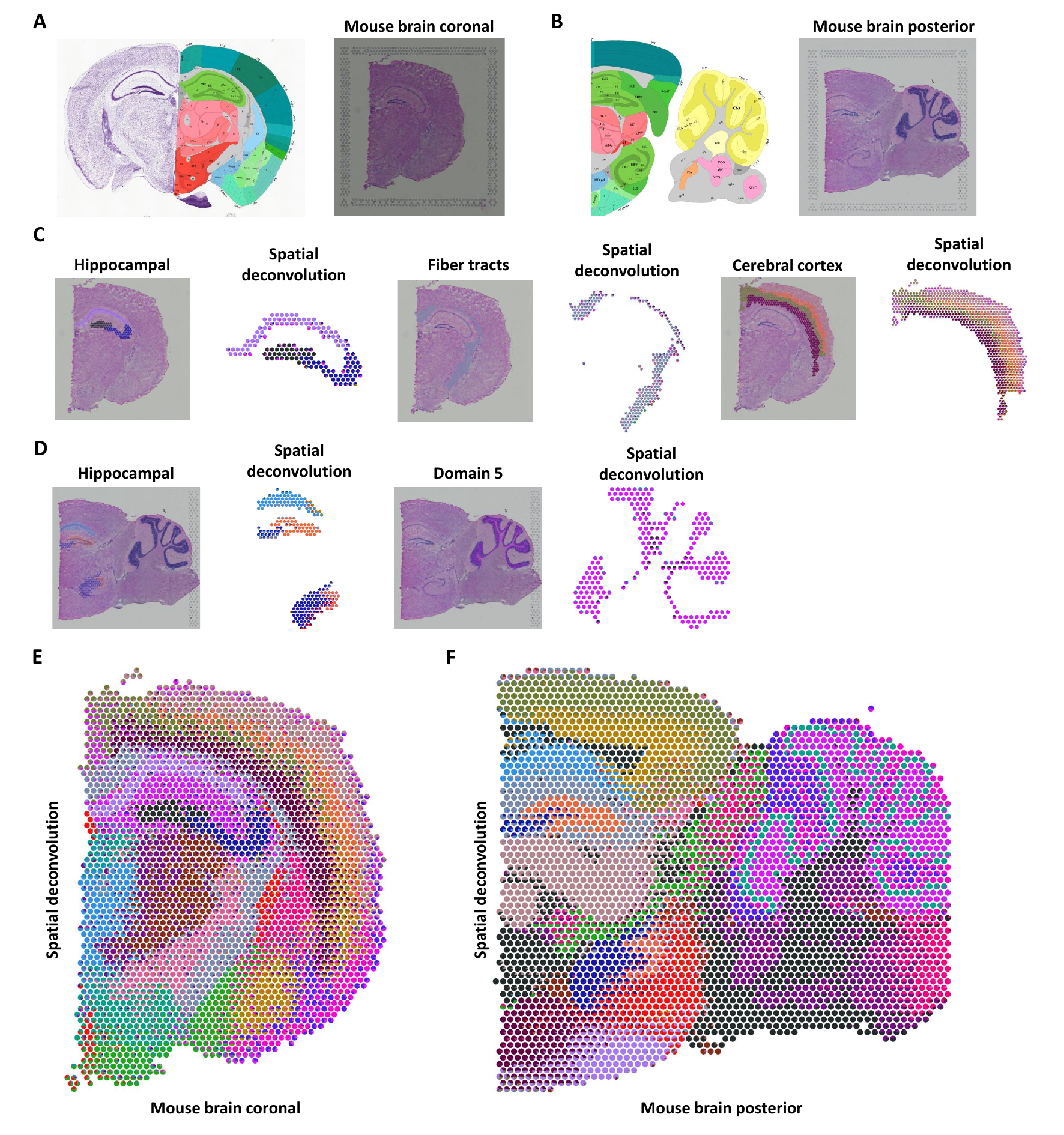}
	\caption{\textbf{Spatial deconvolution of {\MuST}'s results in adult mouse brain section profiled by \href{http://research.libd.org/spatialLIBD/}{10x Visium}. A} Allen Brain Institute reference atlas diagram and H\&E image of the mouse cortex. \textbf{B} Allen Brain Institute reference atlas diagram and H\&E image of the mouse sagittal. \textbf{C} Spatial deconvolution of the identified domains by {\MuST} on coronal mouse brain section. \textbf{D} Spatial deconvolution of the identified domains by {\MuST} on mouse sagittal posterior brain section. \textbf{E} Spatial deconvolution of the coronal mouse brain section. \textbf{F} Spatial deconvolution of the mouse sagittal posterior brain section.}
	\label{fig_Results_Mouse_Brain_Deconvolution}
\end{figure}

\begin{figure}[ht]
	\centering
	\includegraphics[width=\textwidth]{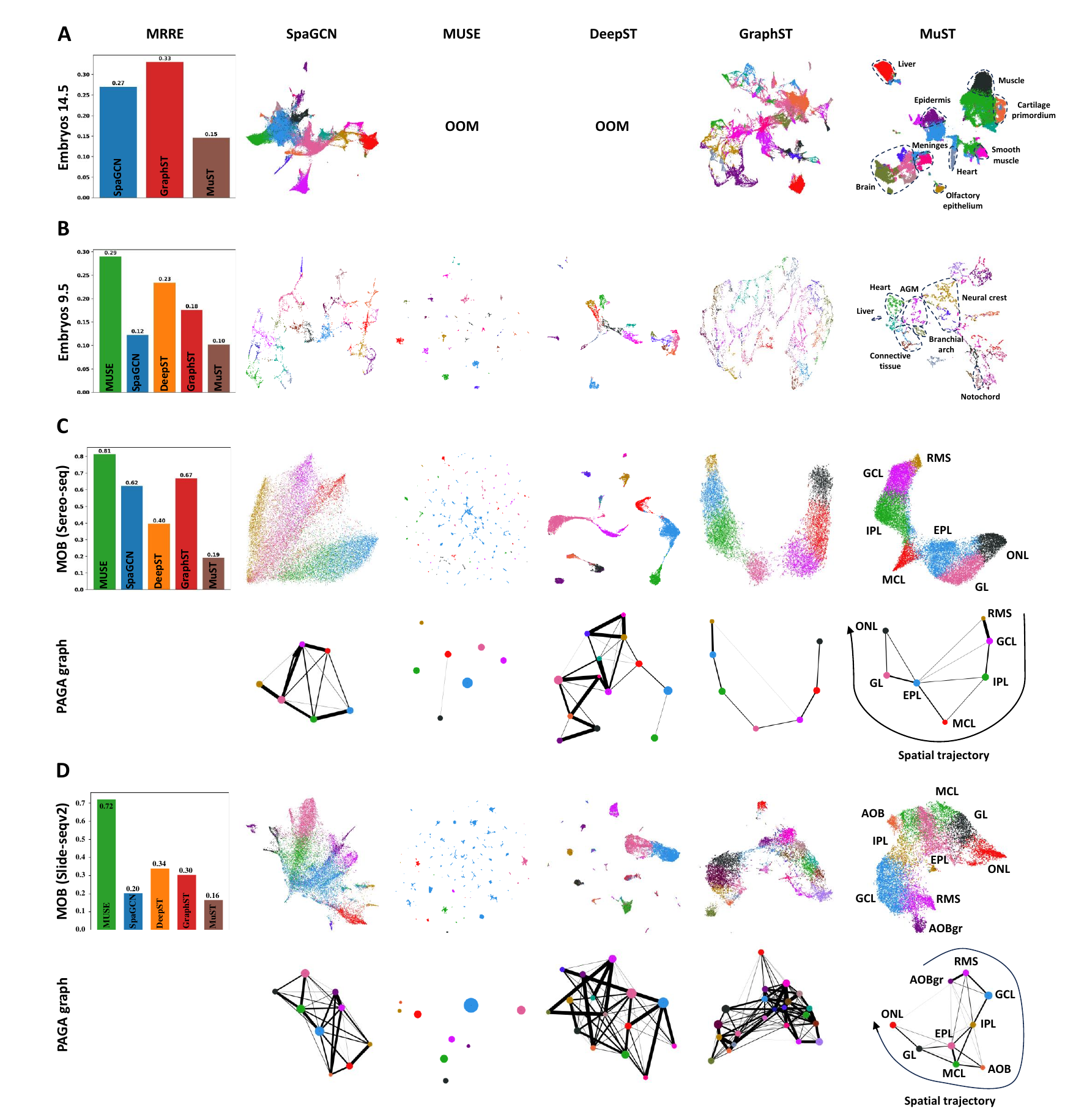}
	\caption{\textbf{Embedding analysis of {\MuST} on the mouse olfactory bulb tissue and mouse embryo. A} Mean relative rank error (MRRE) scores and UMAP visualizations generated by SpcGCN, MUSE, DeepST, GraphST and {\MuST} representations on the Stereo-seq E14.5 mouse embryo data. Among them, methods MUSE and DeepST have an out of memory (OOM) problem. \textbf{B} MRRE scores and spot visualizations generated by the representations of five methods on the Stereo-seq E9.5 mouse embryo data. \textbf{C} MRRE scores, spot visualizations and PAGA graphs generated by the representations of five methods on the Stereo-seq mouse olfactory bulb tissue section. \textbf{D} MRRE scores, spot visualizations and PAGA graphs generated by the representations of five methods on the Slide-seqv2 mouse olfactory bulb tissue section.}
	\label{fig_Results_other_embeddings}
\end{figure}

\begin{figure}[ht]
	\centering
	\includegraphics[width=\textwidth]{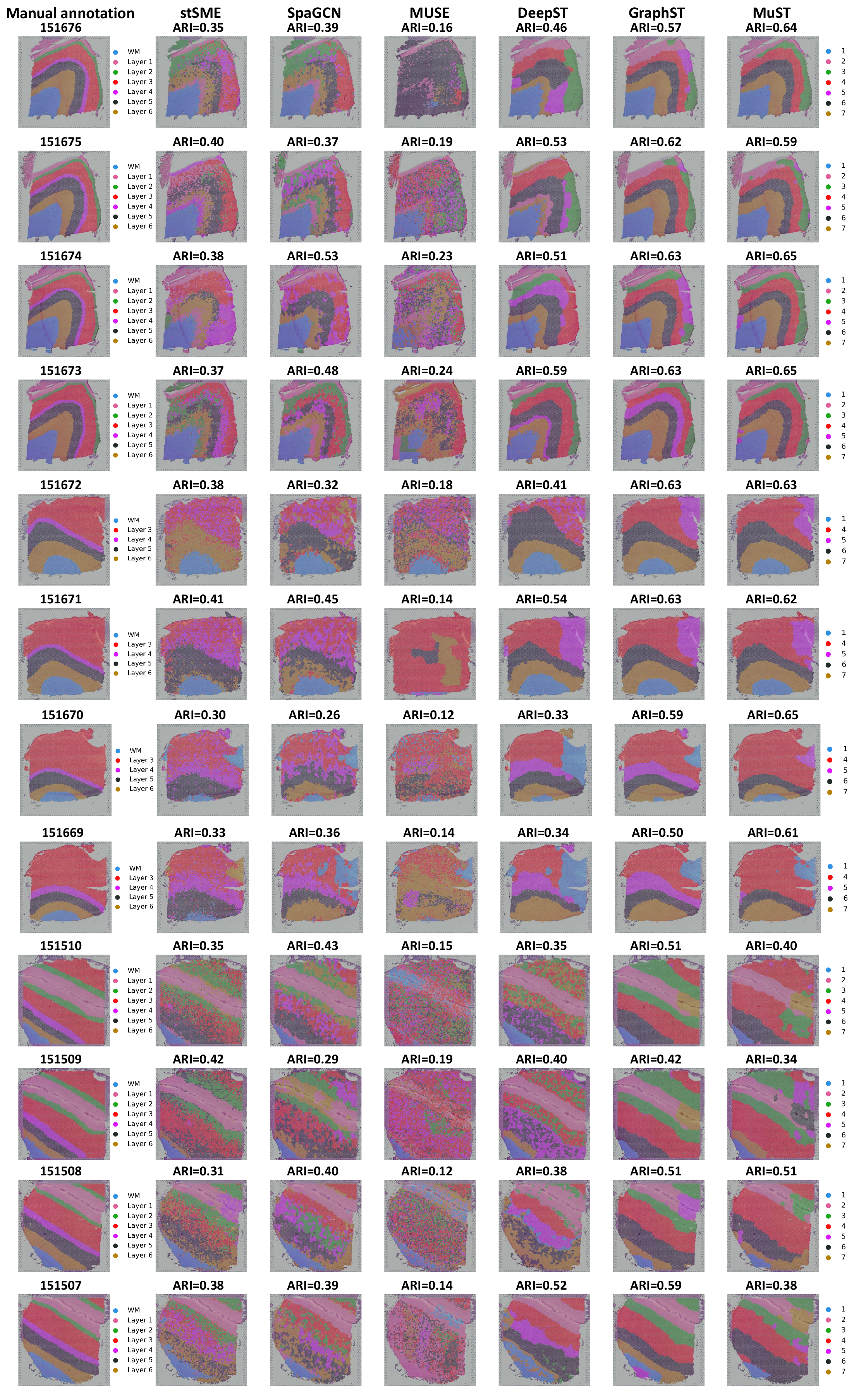}
	\caption{\textbf{Manual annotations and comparison of spatial domains identified by stSME, SpaGCN, MUSE, DeepST, GraphST and {\MuST} on the 12 slices of DLPFC dataset.}}
	\label{fig_Results_DLPFC_all}
\end{figure}

\begin{figure}[ht]
	\centering
	\includegraphics[width=\textwidth]{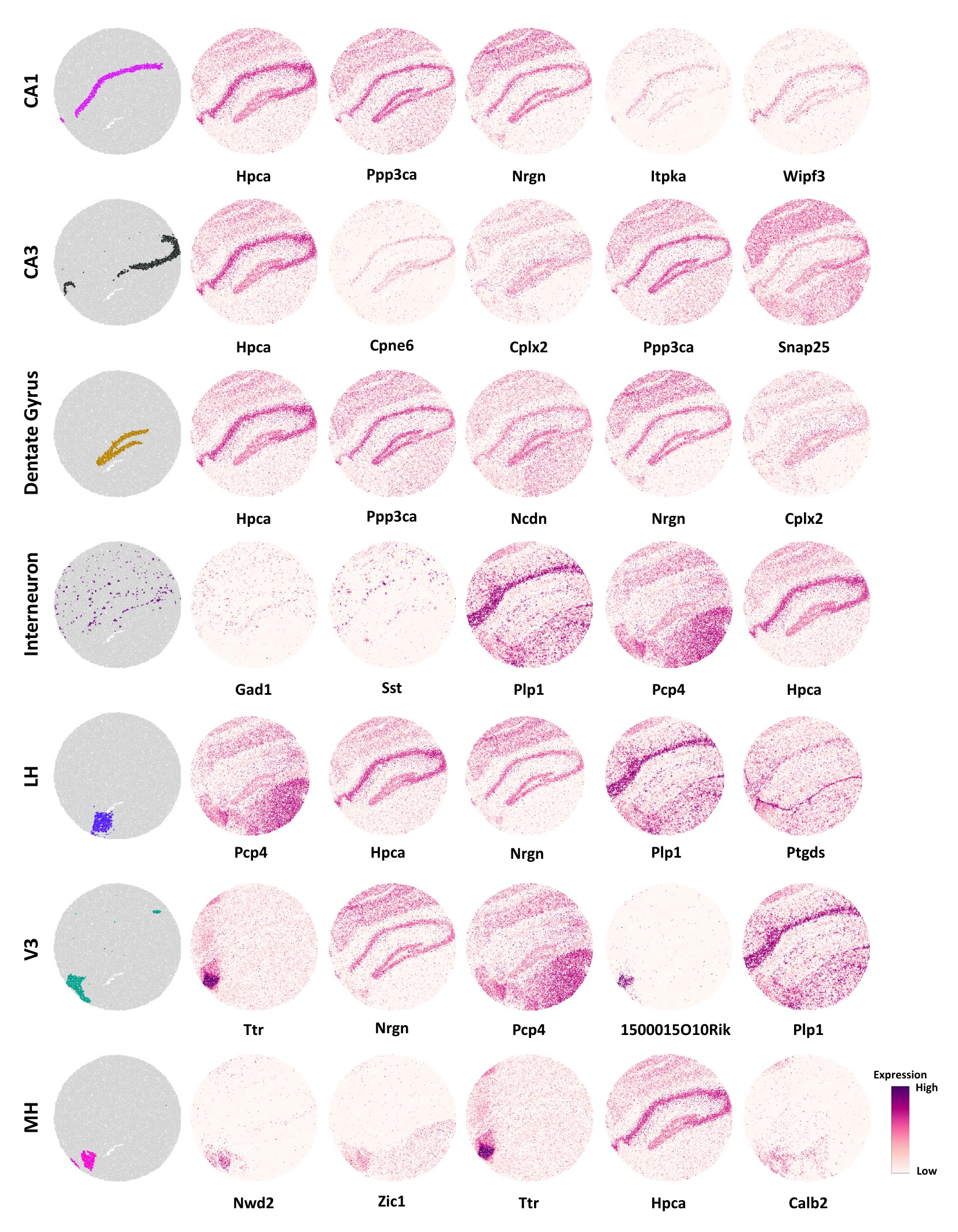}
	\caption{\textbf{Marker genes selection for each anatomical region of mouse hippocampus data acquired with SlideseqV2.}}
	\label{fig_Results_HIP_gene}
\end{figure}

\begin{figure}[ht]
	\centering
	\includegraphics[width=\textwidth]{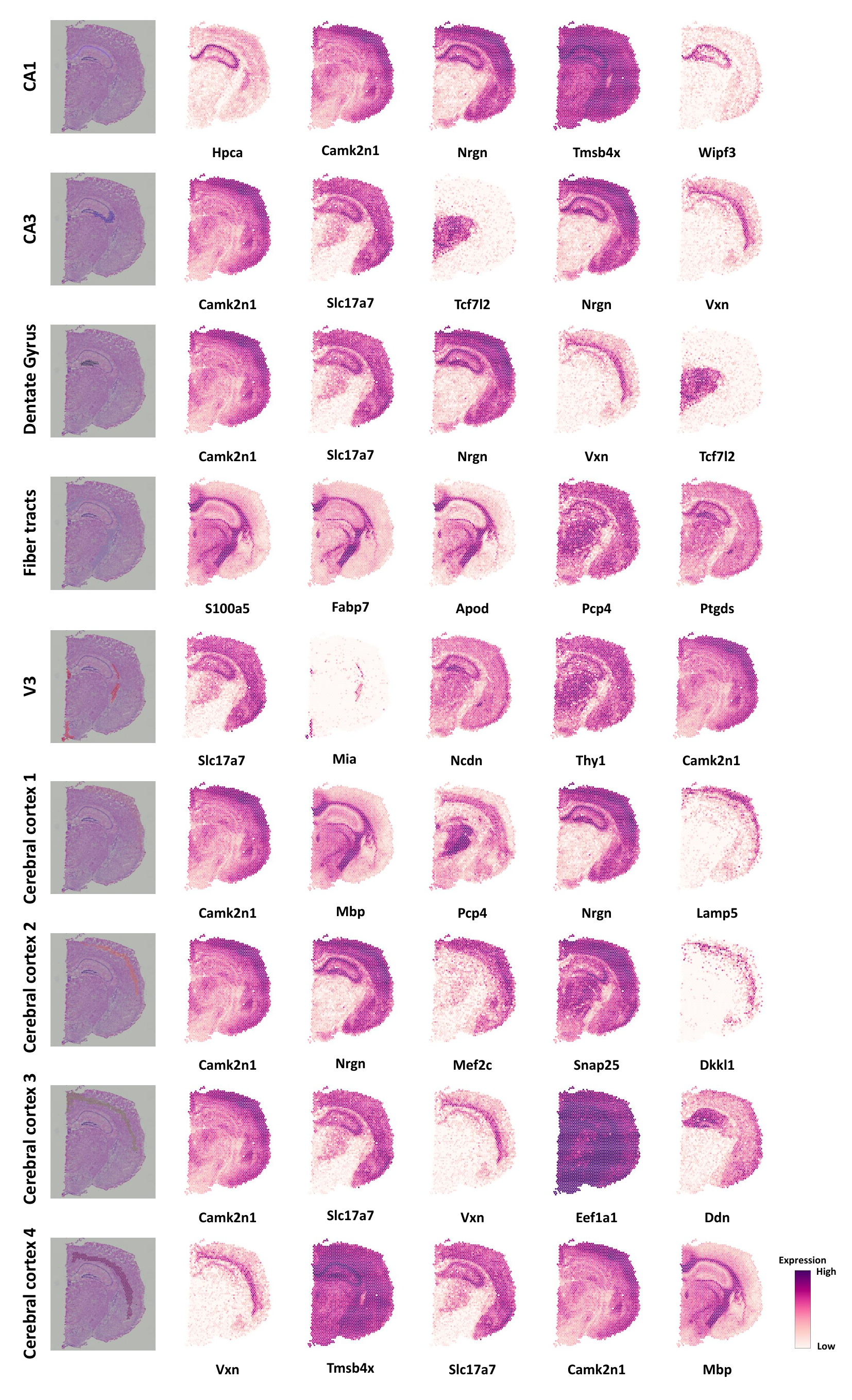}
	\caption{\textbf{Marker genes selection for each region of coronal mouse brain section acquired with 10x Visium.}}
	\label{fig_Results_brain_1_gene}
\end{figure}

\begin{figure}[ht]
	\centering
	\includegraphics[width=\textwidth]{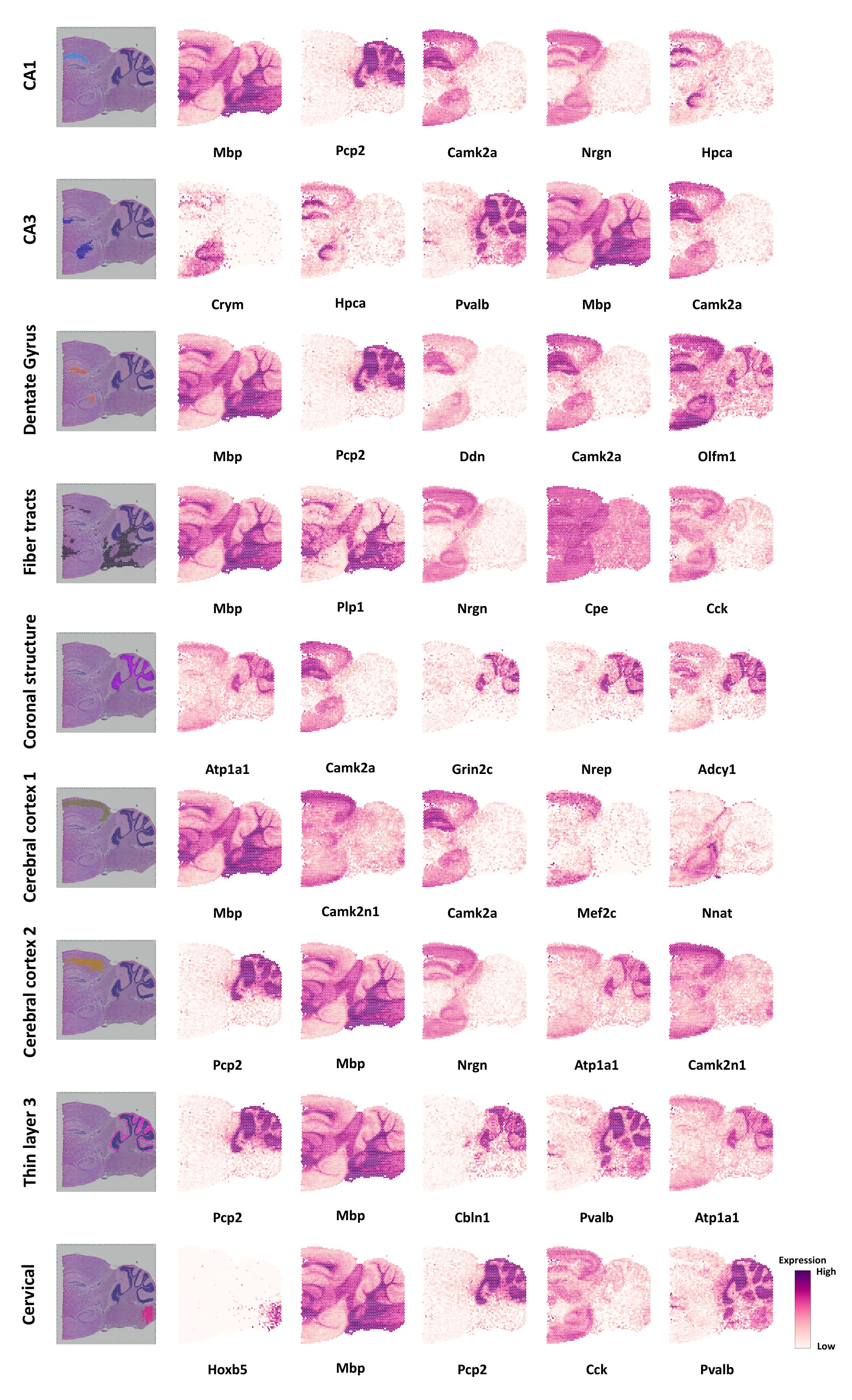}
	\caption{\textbf{Marker genes selection for each region of mouse sagittal posterior brain section acquired with 10x Visium.}}
	\label{fig_Results_brain_2_gene}
\end{figure}

\begin{figure}[ht]
	\centering
	\includegraphics[width=\textwidth]{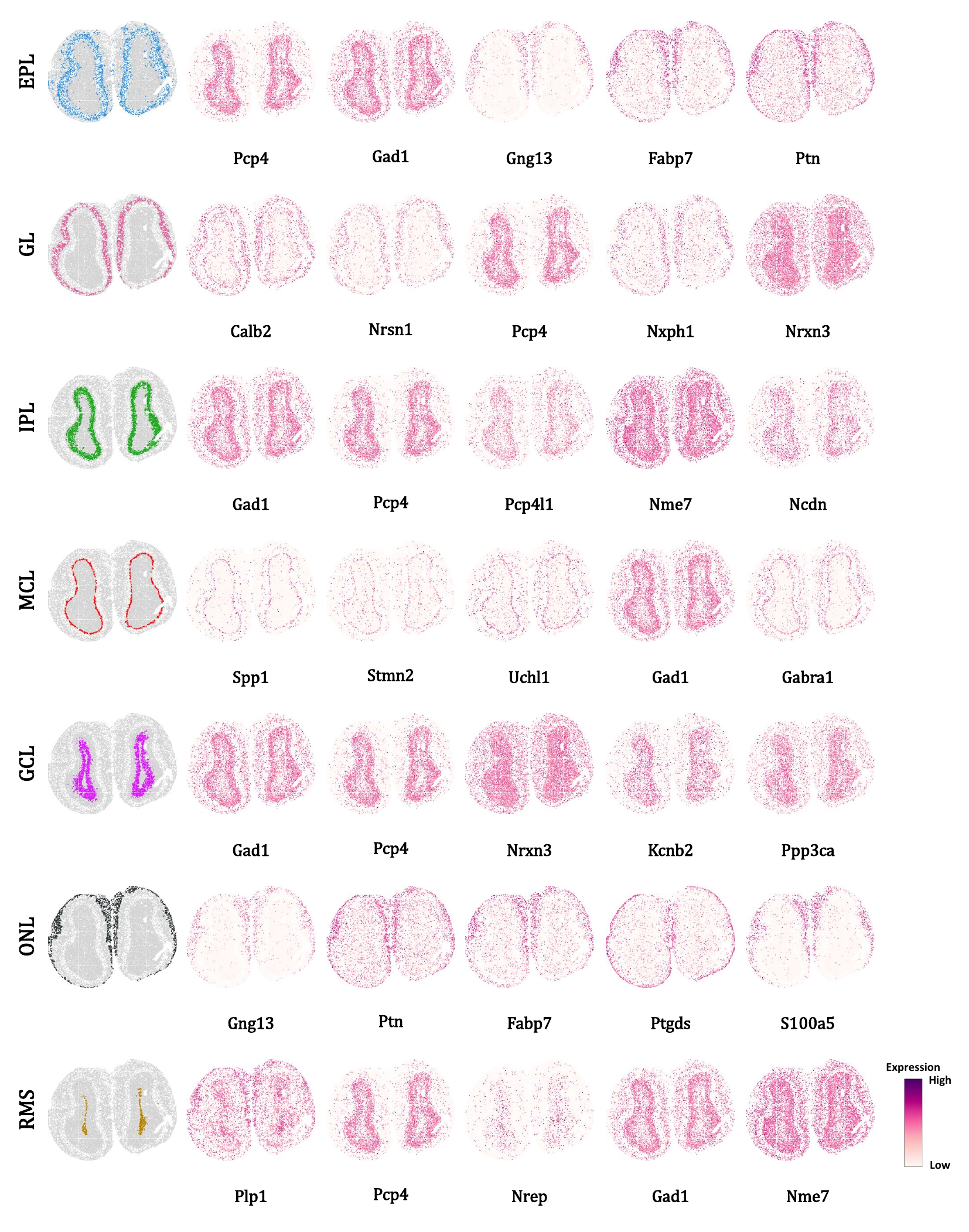}
	\caption{\textbf{Marker genes selection for each laminar organization of coronal mouse olfactory bulb tissue datasets acquired with Stereo-seq.}}
	\label{fig_Results_MOB_Stereo_gene}
\end{figure}

\begin{figure}[ht]
	\centering
	\includegraphics[width=\textwidth]{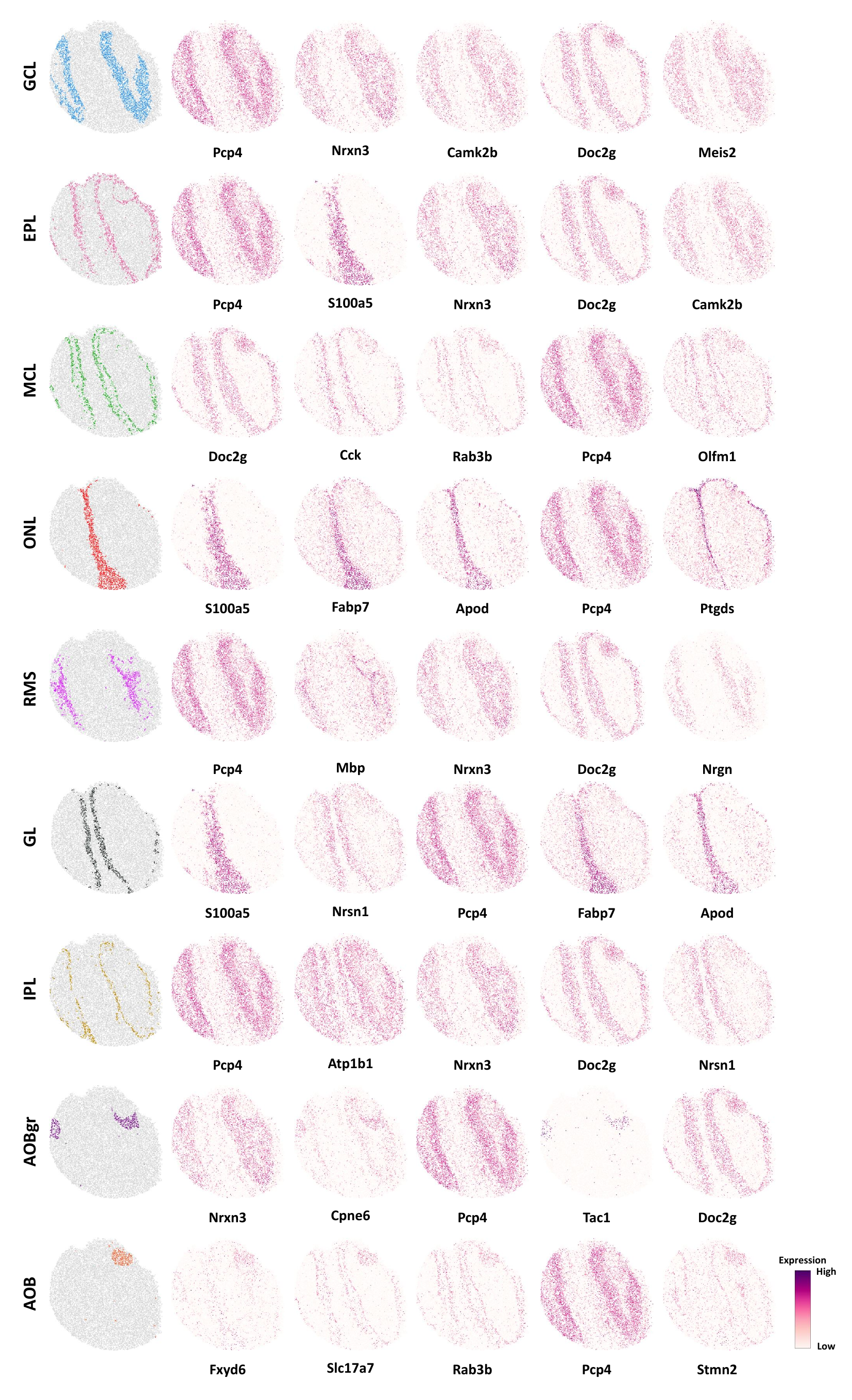}
	\caption{\textbf{Marker genes selection for each laminar organization of coronal mouse olfactory bulb tissue datasets acquired with Slide-seqV2.}}
	\label{fig_Results_MOB_Slide_gene}
\end{figure}

\begin{figure}[ht]
	\centering
	\includegraphics[width=\textwidth]{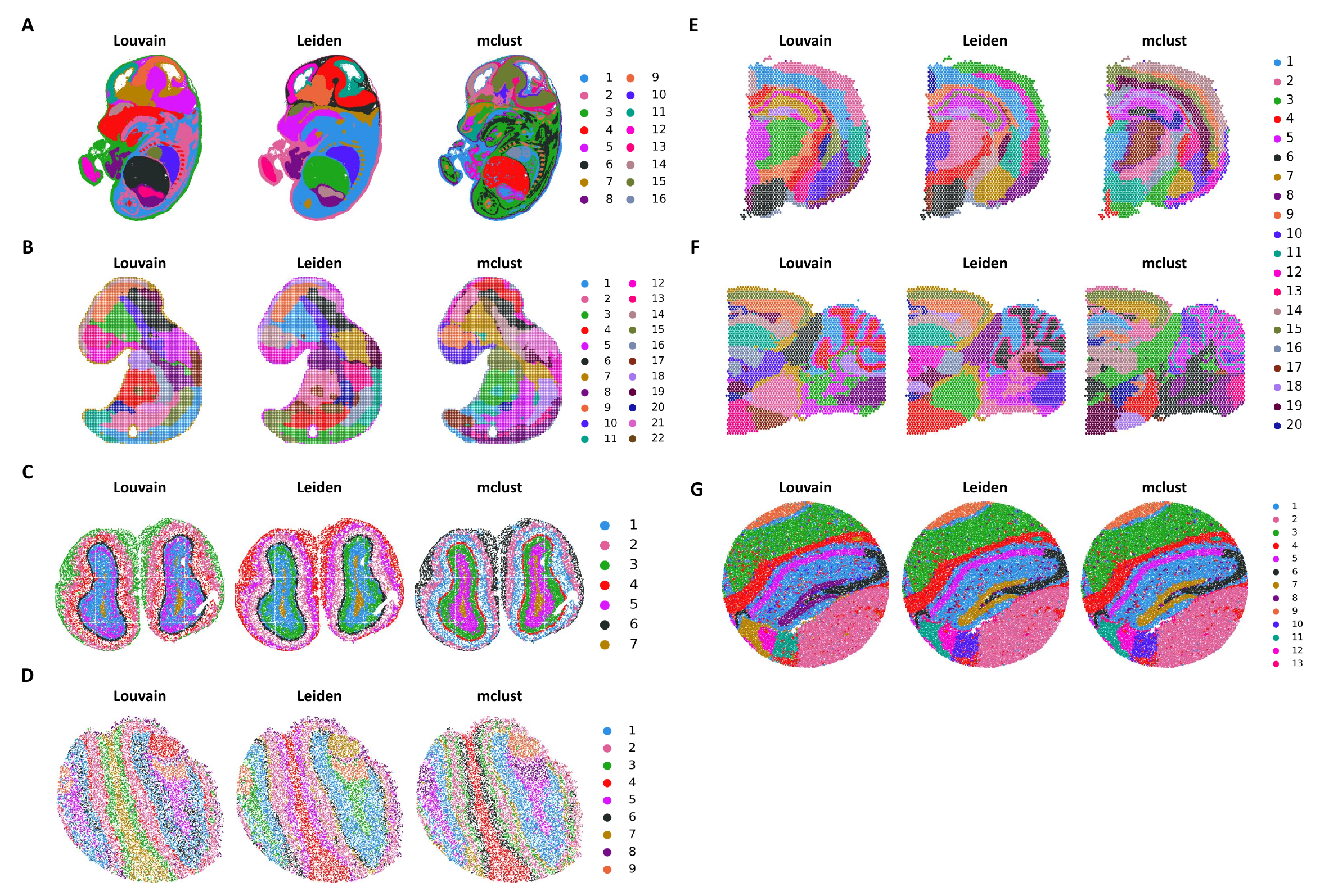}
	\caption{\textbf{Comparison analysis between Leiden, Louvain, and mclust with the output of {\MuST} as input.} Visualization of clustering results from Louvain, Leiden, and mclust on Stereo-seq E14.5 mouse embryo data (A), Stereo-seq 9.5 mouse embryo data (B), Stereo-seq mouse olfactory bulb tissue sections (C), Slide-seqV2 mouse olfactory bulb tissue sections (D), 10x Visium coronal mouse brain section (E), 10x Visium mouse sagittal posterior brain section (F), and SlideseqV2 mouse hippocampus data (G).}
	\label{fig_Results_different_clustering_methods}
\end{figure}
\end{appendices}

\end{document}